\documentclass[twocolumn,preprintnumbers,amsmath,amssymb,nofootinbib,superscriptaddress]{revtex4}

\usepackage{enumerate}
\usepackage{amsmath}
\usepackage{amsfonts}
\usepackage{amssymb}
\usepackage{graphicx, rotating}
\usepackage{epstopdf}
\usepackage{epsfig}
\usepackage{latexsym}
\usepackage{graphicx}
\usepackage{color}
\usepackage{amsmath,amssymb}
\usepackage{slashed}
\usepackage{hyperref}
\hypersetup{colorlinks,citecolor=bluscuro,linkcolor=black,urlcolor=bluscuro}

\definecolor{rossos}{cmyk}{0,1,1,0.55}
\definecolor{bluscuro}{rgb}{0.15, 0.2, .85}
\definecolor{bluchiaro}{cmyk}{1,.3,0.,0.1}
\definecolor{verdechiaro}{rgb}{0.6,.1,0.9}
\definecolor{forestgreen}{rgb}{0.13,0.35,0.13}

\newcommand{\be}{\begin{equation}}
\newcommand{\ee}{\end{equation}}
\newcommand{\bea}{\begin{eqnarray}}
\newcommand{\eea}{\end{eqnarray}}

\def\Im{{\rm Im\,}}

 \hypersetup{colorlinks, citecolor=bluscuro, linkcolor=magenta, urlcolor=verdechiaro}

\begin{document}

\title{\large{REMARKS ON ANALYTICITY AND UNITARITY\\
IN THE PRESENCE OF A STRONGLY INTERACTING LIGHT HIGGS}}
\date{\today}
\author{Alfredo Urbano}  
\email{alfredo.urbano@sissa.it}
\affiliation{SISSA, International School for Advanced Studies, via Bonomea 265, I-34136 Trieste, ITALY.}

\begin{abstract}

Applying  the three axiomatic criteria of Lorentz invariance, analyticity and unitarity
to scattering amplitudes involving the Goldstone bosons and the Higgs boson,
 we derive a general sum rule for the Strongly Interacting Light Higgs Lagrangian.
  This sum rule connects the IR coefficient $c_H$ to the UV properties of the theory, 
  and can be used, for instance, to capture the role of resonances in processes
  like $V_{\rm L}V_{\rm L}\to hh$ and $V_{\rm L}V_{\rm L}\to 
  V_{\rm L}V_{\rm L}$, with $V=W^{\pm},Z$. 

\end{abstract}

\maketitle
 
\section{Introduction}

The Higgs boson \cite{Higgs,Higgs2,Higgs3} was found in July 2012 after a near half-century search \cite{Aad:2012tfa,Chatrchyan:2012ufa}. As predicted by the Standard Model (SM), the Higgs boson is a scalar particle, and all the experimental data collected so far at the LHC seem to point towards a positive parity \cite{ATLASwiki,CMSwiki}. The value of the mass is $m_h={125.5\pm 0.2_{\rm stat}}{^{+0.5}_{-0.4}}_{\rm sys}$ GeV measured by ATLAS \cite{ATLAS_Mass}, and $m_h=125.7 \pm 0.3_{\rm stat}\pm 0.3_{\rm sys}$ GeV measured by CMS \cite{CMS_Mass}.
 Despite this astonishing discovery, however, none of the mysteries related to the existence of the Higgs boson have been solved. Light scalars are unnatural in quantum field theory, unless a specific mechanism keeps their mass safe from large radiative corrections. The most elegant way to solve this problem is to protect the Higgs boson using a symmetry. This particular theoretical framework is realized in the context of  Composite Higgs models, where the Higgs boson arises as pseudo Nambu-Goldstone boson of a spontaneously broken global symmetry \cite{Kaplan:1983fs,Georgi:1984af,Kaplan:1983sm,Dugan:1984hq,Contino:2003ve,Agashe:2004rs}.\footnote{A more precise formulation of the hierarchy problem in the context of Composite Higgs models is the following. On a general ground, it is well known that the Higgs mass -- via quantum corrections -- is quadratically sensitive to the existence of  any new physics beyond the SM, i.e. $\Delta m_h^2 \sim (g^2/16\pi^2)\Lambda^2$ where $g$ is a coupling constant and $\Lambda$ is the scale of new physics.  The so-called ``big hierarchy problem'' refers to the question why is $m_h \ll \Lambda$. The most straightforward solution to this problem relies on the following arguments. First,
 the scale of new physics has to be relatively low, $\Lambda \lesssim$ TeV; second, the new physics, irrespective from its nature, has to render the Higgs boson insensitive to further quantum corrections above the TeV scale. In Composite Higgs model the occurrence of the latter condition is ensured by compositeness itself -- i.e. by the fact that in the fundamental theory above the TeV scale there exists no scalar operators of dimension less than 4 that can be added 
 to the Lagrangian -- while the condition $m_h\ll $ TeV follows from the pseudo Nambu-Goldstone nature of the Higgs scalar. The ``little hierarchy problem'', on the contrary, refers to the lack of direct and/or indirect evidences of this new physics at the TeV scale. In Composite Higgs models this problem is mitigated by the fact that
 the Higgs mass correction is actually of the form $\Delta m_h^2 \sim (g^2/16\pi^2)f^2$ and is set by the global symmetry breaking scale $f$ rather by the new physics scale $\Lambda$. Following the previous discussion, as a consequence, one would expect $f\lesssim$ TeV while the cut-off scale of the theory will lie at higher values $\Lambda\sim 4\pi f$, thus alleviating the little hierarchy problem.}
This scenario has profound phenomenological implications; it predicts potentially large deviations in the couplings of the Higgs boson with the SM gauge bosons and fermions, as well as the existence of new resonances, with a mass around the TeV scale, coupled to the Higgs doublet. Unambiguous fingerprints of compositeness, therefore, could be present in sizable deformations of the Higgs couplings, and 
significantly enhanced cross sections describing scattering processes between the Higgs boson and/or the longitudinal  gauge bosons $W_{\rm L}^{\pm},Z_{\rm L}$. The former are under scrutiny at the LHC \cite{Aad:2013wqa,CMS}, and the current experimental bounds are still compatible with the presence of deviations from the SM predictions, especially considering loop-induced couplings (see, for instance, Refs.~\cite{Falkowski:2013dza,Giardino:2013bma,Pomarol:2013zra}). As far as the latter is concerned,  processes like $V_{\rm L}V_{\rm L}\to hh$, $V=W^{\pm},Z$ have a distinctive signature at the LHC: 
the production of two Higgses in association with two forward jets, well separated in pseudorapidity, related to the primary partons that radiate the $V_{\rm L}V_{\rm L}$ pair. The possibility to detect these processes at the LHC is extremely challenging, given the tiny value of the corresponding SM cross sections \cite{Contino:2010mh,Baglio:2012np}; on the other hand, this also implies that new physics effects -- in particular due to the  s-channel exchange of a new resonance in $V_{\rm L}V_{\rm L}\to hh$, $V_{\rm L}V_{\rm L}\to V_{\rm L}V_{\rm L}$  -- are more likely to be seen \cite{Contino:2011np}.

Actually, apart from experimental complications, there exists also a nontrivial theoretical obstruction. Composite Higgs models postulate the existence of a new strongly-coupled sector to which these resonances belong,  thus making the usual perturbative approach completely useless. The possibility to make model-independent predictions without any knowledge of the underlying UV-completion of the theory may seem, as a consequence, completely hopeless. 

In the sixties, the ambitious goal of the ``S-matrix theory'' was to compute the elements of the S-matrix by requiring them to respect three general properties that ought to be valid independently of the actual existence of a Lagrangian description: Lorentz invariance, analyticity and unitarity  \cite{Gribov:2009zz}. 

The S-matrix theory was developed in order to describe the transition amplitudes in the presence of the strong interaction responsible for nuclear forces, like for instance the pion-nucleon scattering. 
\textit{Mutatis mutandis}, we can try to apply the same basic principles in the presence of the strong dynamics of a Composite Higgs model. In this way, one can pursue the possibility to study the structure of 
scattering amplitudes through an elegant union between the analytical approach and the exploitation of the underlying symmetries.  

Keeping this aim in mind,  in this paper we apply the principles of the S-matrix theory to study the scattering amplitudes in presence of a Strongly Interacting Light Higgs (SILH), and in particular we focus on the processes involving the Goldstone bosons and the Higgs boson. Thanks to the Equivalence Theorem \cite{Chanowitz:1985hj}, in fact, at high energies these processes formally take the place 
of $V_{\rm L}V_{\rm L}\to hh$, $V_{\rm L}V_{\rm L}\to V_{\rm L}V_{\rm L}$. 

This work is organized as follows. In Section~\ref{sec:SILH} we describe the theoretical setup of our computation, provided by the SILH effective Lagrangian \cite{Giudice:2007fh}. In Section~\ref{sec:SumRule}, using a dispersion relation, we derive a sum rule that is the main result of this paper. In Section~\ref{sec:Discussion} we discuss some phenomenological implications. Finally, we conclude in Section~\ref{sec:Conclusion}. In Appendix~\ref{app:Smatrix} we summarize the basic principles of the S-matrix theory. In Appendix~\ref{eq:hhScattering} we 
construct the scattering amplitudes used in Section~\ref{sec:SumRule}. In Appendix~\ref{App:FMBound} we generalize the Froissart-Martin bound to inelastic scattering amplitudes. In Appendix~\ref{app:Non-compact} we construct the non-linear $\sigma$-model Lagrangian describing the coset $SO(4,1)/SO(4)$.

\section{Setup: The SILH effective Lagrangian}\label{sec:SILH}

The scalar sector of the SM is described by the following Lagrangian
\begin{equation}\label{eq:HiggsSectorSM}
\mathcal{L}_{H}=(D_{\mu}H)^{\dag}(D^{\mu}H)-\mu^2|H|^2-\frac{\lambda}{2}|H|^4~,
\end{equation}
with $\mu^2 <0$; $H$ is the usual Higgs doublet
\begin{equation}\label{eq:FundamentalDoublet}
H=
\left(
\begin{array}{c}
   \pi^+ \\
     \frac{1}{\sqrt{2}}(h+i\pi^0)
\end{array}
\right)
\end{equation}
 with vacuum expectation value (vev) 
$\langle H\rangle = (0, v)^T/\sqrt{2}$, and 
\begin{equation}
D^{\mu}H =
\partial_{\mu}H + \frac{ig_{\rm L}}{2}\sigma^{a}W^{a}_{\mu}H + \frac{ig_{\rm Y}}{2}B_{\mu}H~,
\end{equation}
is the covariant derivative related to the gauging $SU(2)_{\rm L}\otimes U(1)_{\rm Y}$, 
being $\sigma^{a=1,2,3}$
the usual Pauli matrices. In Eq.~(\ref{eq:FundamentalDoublet}) $\pi^{\pm}\equiv(\pi^1\mp i\pi^2)/\sqrt{2}$, $\pi^0$ are
the Goldstone bosons while $h$ is the Higgs boson.
  The minimum of the potential occurs for $v^2 = -2\mu^2/\lambda$, and after electroweak symmetry breaking the Higgs boson acquires the mass $m_h^2 = \lambda v^2$.

The Higgs Lagrangian $\mathcal{L}_{H}$ possesses, in the limit $g_{\rm Y}\to 0$, the larger global symmetry
$SO(4)\approx SU(2)_{\rm L}\otimes SU(2)_{\rm R}$, spontaneously broken by the Higgs vev into the diagonal custodial subgroup 
$SO(3)\approx SU(2)_{\rm C}$.
On the one hand -- in the unbroken phase -- the Goldstone bosons and the Higgs boson  
transform under the action of the global symmetry $SO(4)$ according to its fundamental representation or, equivalently, 
according to the bi-doublet representation of $SU(2)_{\rm L}\otimes SU(2)_{\rm R}$;
 on the other one -- after electroweak symmetry breaking -- under the action of the custodial group the Higgs boson transforms as a singlet, 
while the Goldstone bosons transform as a triplet, parametrizing the coset $SU(2)_{\rm L}\otimes SU(2)_{\rm R}
/SU(2)_{\rm C}$.

Given this setup, 
one may wonder if the global symmetry $SO(4)\approx SU(2)_{\rm L}\otimes SU(2)_{\rm R}$
 is just an accidental property encoded in the Higgs Lagrangian of the SM or if its presence is rooted in
 a more profound theoretical ground. 
 
 The latter scenario is realized  in the context of Composite Higgs models, in which the Higgs is a pseudo Nambu-Goldstone boson, and -- in analogy with the pions in QCD --  it originates from the spontaneous breaking of a global symmetry. 
In more detail, the picture to bear in mind is the following.  In addition to the elementary sector, formed by all the SM fields with the exception of the Higgs doublet,\footnote{For simplicity, we do not mention in this brief discussion the possibility that also the top quark might belong to the strong sector \cite{Giudice:2007fh}.} there exists a composite sector -- around the TeV scale -- described by a new fundamental strongly-coupled theory, and characterized by the global symmetry $\mathcal{G}$.
At some new scale $f$,  this global symmetry is spontaneously  
 broken by a dynamical condensate into the subgroup $\mathcal{H}\supset SU(2)_{\rm L}\otimes U(1)_{\rm Y}$. The crucial assumption is that, in the limit in which all the SM couplings are zero, the Higgs doublet $H$ is an exact Nambu-Goldstone boson doublet  living in the coset $\mathcal{G}/\mathcal{H}$.  Assuming that the strong sector preserves the custodial symmetry,
 the minimal choice turns out to be $\mathcal{G}/\mathcal{H}=SO(5)/SO(4)$. The picture is completed by the SM gauge and Yukawa
  couplings; they break explicitly the global symmetry, thus making the Higgs a pseudo Nambu-Goldstone boson, and generating radiatively the electroweak potential.

  The general features of this framework, and in particular the predicted deviations from the SM, can be captured in a model-independent way by
   using the SILH effective Lagrangian \cite{Giudice:2007fh}. In this paper we are interested 
   in the operators of the SILH Lagrangian that involve only the Higgs doublet, and therefore -- at dim-$6$ -- 
   we have
 \begin{eqnarray}
 \mathcal{O}_{H}&\equiv& \frac{c_{H}}{2f^2}\partial_{\mu}(H^{\dag}H)\partial^{\mu}(H^{\dag}H)~,\\
  \mathcal{O}_{6}&\equiv& -\frac{c_6 \lambda}{f^2} (H^{\dag}H)^3~.
 \end{eqnarray} 
  Notice that in the following, in order to simplify the notation, 
   we shall refer to the generic component 
   of the doublet $H$ using the symbol $\pi^a$, i.e. $\pi^a=\pi^{\pm},\pi^0,h$.  
   Furthermore, we focus only on the scattering processes
   $\pi^a\pi^b\to \pi^c\pi^d$  whose amplitude grows with the energy.
As a consequence, we concentrate on the derivative of the Goldstone 
   doublet described by the operator $\mathcal{O}_{H}$.  In the next Section 
   we shall derive a general sum rule for the SILH Lagrangian studying the scattering processes
   $\pi^a\pi^b\to \pi^c\pi^d$. The underlying assumption is that the UV-completion of the SILH Lagrangian respects the postulates of Lorentz invariance, analyticity and unitarity (see Ref.~\cite{Gribov:2009zz}, and Appendix~\ref{app:Smatrix} for a review of the basic definitions). This is a fundamental requirement that we expect to be true in any string-inspired UV-completion.\footnote{See Ref.~\cite{Caracciolo:2012je} for a recent discussion about the UV-completion of Composite Higgs
models with partial compositeness.}
  
 \section{Analyticity and Unitarity: IR/UV connection}\label{sec:SumRule}

\begin{figure}[!htb!]
  \includegraphics[width=1 \linewidth]{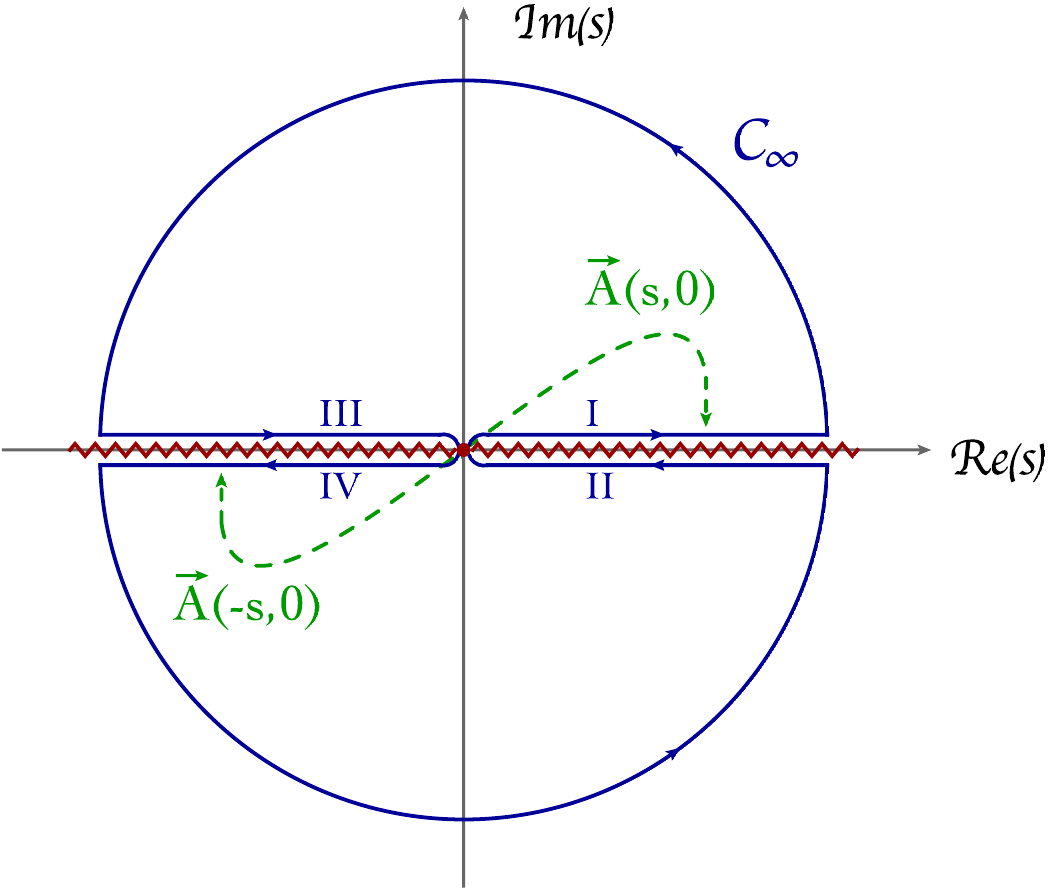}
 \caption{\textit{Contour of integration $\mathcal{C}$ (counterclockwise, blue solid line) in the
 complex s-plane according to Eq.~(\ref{eq:MasterIntegral}), decomposed into
 the four contributions (I-IV) surrounding the cuts (red zigzag line), and the contribution from the big circle 
 at infinity $\mathcal{C}_{\infty}$. The green dashed line pictorially represents the crossing transformation relating
 the s- and u-channel amplitudes in the forward limit 
 [see Eqs.~(\ref{eq:CUTS},\,\ref{eq:MasterCrossing})  and Appendix~\ref{App:Crossing}].
 }}\label{fig:Contour}
\end{figure}

  Classifying the Goldstone bosons $\pi^{\pm}, \pi^0$ and the Higgs boson $h$ according to the representation
    $(\textbf{2},\textbf{2})$ of $SU(2)_{\rm L}\otimes SU(2)_{\rm R}$, it is straightforward to realize that the scattering $\pi\otimes \pi^{\prime}$
    has the following structure 
    \begin{equation}\label{eq:SUDecomposition1}
(\textbf{2},\textbf{2})\otimes (\textbf{2},\textbf{2})
=(\textbf{1},\textbf{1})
\oplus (\textbf{3},\textbf{1})\oplus (\textbf{1},\textbf{3})\oplus (\textbf{3},\textbf{3})~.
\end{equation}
This simply means, as a consequence, that the scattering amplitude
 $\mathcal{A}_{\pi^a \pi^b\to \pi^c \pi^d}(s,t)$ 
 describing the generic process
 \begin{equation}
\pi^a(p_1) + \pi^b(p_2) \to \pi^c(p_3) + \pi^d(p_4)~,
\end{equation} 
with $s=(p_1+p_2)^2$, $t=(p_1 - p_3)^2$, 
 can always be decomposed in terms of its projections
with definite $SU(2)_{\rm L}\otimes SU(2)_{\rm R}$ quantum numbers. In full generality,
therefore, instead of a single specific  amplitude we focus on the combination
\begin{equation}\label{eq:Projections}
\mathcal{A}(s,t=0)\equiv \sum_{{\rm IJ}=00,10,01,11} \kappa_{{\rm IJ}}~\mathcal{A}_{{\rm IJ}}(s,t=0)~,
\end{equation}
where $ \kappa_{{\rm IJ}}$ are arbitrary constants, and $\mathcal{A}_{{\rm IJ}}(s,t)$ are the 
projected amplitudes according to the decomposition in Eq.~(\ref{eq:SUDecomposition1}). 
In Eq.~(\ref{eq:Projections}), moreover, we have explicitly considered the forward limit $t=0$. In Appendix~\ref{eq:hhScattering}
we compute in detail all  the scattering amplitudes
  $\mathcal{A}_{\pi^a \pi^b\to \pi^c \pi^d}(s,t)$ as a function of the projections 
$\mathcal{A}_{{\rm IJ}}(s,t)$ [see Eqs.~(\ref{eq:Amp1}-\ref{eq:Amp7})].
Following Refs.~\cite{Adams:2006sv,Low:2009di,Nicolis:2009qm,Falkowski:2012vh}
we compute the  integral
\begin{equation}\label{eq:MasterIntegral}
\mathcal{I}=\int_{\mathcal{C}}\frac{ds}{2\pi i}\frac{\mathcal{A}(s,t=0)}{s^2}~,
\end{equation}
 where the contour of integration $\mathcal{C}$ is displayed 
in Fig.~\ref{fig:Contour}, and can be decomposed into two contributions: 
the contribution from the parts (denoted as I-IV in Fig.~\ref{fig:Contour}) surrounding 
the unitarity cuts,\footnote{The contour $\mathcal{C}$ lies on the first Riemann sheet (the physical sheet), where the only singularities of a scattering amplitude are simple poles and branch cuts. Poles associated with resonances, on the contrary, lie on the second Riemann sheet, and they play no role in the computation of the integral in Eq.~(\ref{eq:MasterIntegral}).} and the contribution from the big circle at infinity, $\mathcal{C}_{\infty}$.\footnote{See also Refs.~\cite{Distler:2006if,Manohar:2008tc} for the computation of similar integrals in the context of the longitudinal $W_{\rm L}W_{\rm L}$ scattering, and
Refs.~  \cite{QCDsumRule,Adler:1968hc,Ecker:1988te,Knecht:1995tr,Bijnens:1997vq,Cirigliano:2006hb,Nieves:2011gb,Greynat:2013zsa,Ananthanarayan:2000ht,Ananthanarayan:1994hf,Ananthanarayan:2001uy,Ananthanarayan:2000cp,Ananthanarayan:1997yi} for related studies in QCD.}
Notice that the scattering amplitude $\mathcal{A}(s,t=0)$ in Eq.~(\ref{eq:MasterIntegral}) has been promoted to an analytic function of the complex variable $s$ defined in the complex plane.
The integral $\mathcal{I}$ can be computed in two different ways, providing a connection between the IR and UV 
behavior of the theory. 

$1.$~We compute the residual value of $\mathcal{I}$ at $s=0$, where the scattering amplitudes $\mathcal{A}_{\pi^a \pi^b\to \pi^c \pi^d}(s,t)$ can be written explicitly extracting the interactions encoded in the operator $\mathcal{O}_H$.
This approach, relying on the lowest order of the effective field theory description provided by the SILH Lagrangian,  
captures the IR limit of the theory. By direct computation we find \cite{Giudice:2007fh,Low:2009di}
 \begin{eqnarray}
  \mathcal{A}_{\pi^{\pm}\pi^{\mp}\to \pi^{0}\pi^{0}}(s,t) &=& \frac{c_H s}{f^2}~,\\
    \mathcal{A}_{\pi^{\pm}\pi^{0}\to \pi^{\pm}\pi^{0}}(s,t) &=& \frac{c_H t}{f^2}~,\\
     -\mathcal{A}_{\pi^{\pm}\pi^{\pm}\to
 \pi^{\pm}\pi^{\pm}}(s,t) &=& \frac{c_H s}{f^2}~,\\
  \mathcal{A}_{\pi^{\pm}\pi^{\mp}\to
 \pi^{\pm}\pi^{\mp}}(s,t) &=& \frac{c_H(s+t)}{f^2}~,
 \end{eqnarray}  
with equal amplitudes obtained substituting 
$\pi^{0}$ with $h$, i.e. $ \mathcal{A}_{\pi^{\pm}\pi^{\mp}\to \pi^{0}\pi^{0}}(s,t)=
 \mathcal{A}_{\pi^{\pm}\pi^{\mp}\to hh}(s,t)$. Evaluating the  corresponding projections according to
 Eqs.~(\ref{eq:A11}-\ref{eq:A01}), and taking the forward limit we obtain 
 \begin{equation}
 \mathcal{A}(s,t=0) \stackrel{{\rm IR}}{=} \frac{c_H s}{f^2}(3\kappa_{00} + \kappa_{10}
 +\kappa_{01} -\kappa_{11})~,
 \end{equation}
 and, as a consequence 
 \begin{equation}\label{eq:MASTER1}
 \mathcal{I} =  \frac{c_H}{f^2}(3\kappa_{00} + \kappa_{10}
 +\kappa_{01} -\kappa_{11})~.
 \end{equation}

$2.$~The second method is based on the explicit computation of the integral following the contour $\mathcal{C}$
\begin{equation}\label{eq:Separation}
\mathcal{I}=\int_{{\rm I-IV}}\frac{ds}{2\pi i}\frac{\mathcal{A}(s,t=0)}{s^2} +
\int_{\mathcal{C}_{\infty}}\frac{ds}{2\pi i}\frac{\mathcal{A}(s,t=0)}{s^2}~,
\end{equation}
in which we have separated the contribution from the cuts and the contribution from the big circle at infinity. 
Let us consider first the contribution from the cuts; dropping the t-dependence we have
\begin{eqnarray}\label{eq:CUTS}
&&\int_{{\rm I-IV}}\frac{ds}{2\pi i}\frac{\mathcal{A}(s)}{s^2} =\nonumber \\ &&
\int_{0}^{\infty}\frac{ds}{2\pi i}\lim_{\epsilon\to 0^+}\left[
\frac{\mathcal{A}(s+i\epsilon)-\mathcal{A}(s-i\epsilon)}{s^2}
\right]-\nonumber \\
&&\int_{0}^{\infty}\frac{ds}{2\pi i}\lim_{\epsilon\to 0^+}\left[
\frac{\mathcal{A}(-s-i\epsilon)-\mathcal{A}(-s+i\epsilon)}{s^2}
\right]~,\nonumber\\
\end{eqnarray}
where the first (second) term represents the discontinuity of the scattering amplitude across the right (left) cut (see Fig.~\ref{fig:Contour}). 
As customary in this kind of computation, analyticity allows us
to apply  the crossing symmetry transformation that 
relates the amplitude in the u-channel and the amplitude in the s-channel; 
in terms of the projection $\mathcal{A}_{\rm IJ}$ in Eq.~(\ref{eq:Projections}) we have
the following matrix equation
\begin{equation}\label{eq:MasterCrossing}
\vec{\mathcal{A}}(-s)=C_{su}^{-1}~\vec{\mathcal{A}}(s)~,
\end{equation}
where 
$\vec{\mathcal{A}}\equiv (\mathcal{A}_{00},
\mathcal{A}_{10},\mathcal{A}_{01},\mathcal{A}_{11})^T$. In Appendix~\ref{App:Crossing}
we compute explicitly this transformation, and the matrix $C_{su}$ is given in Eq.~(\ref{eq:CrossingMatrixU}).
All in all, using the decomposition in Eq.~(\ref{eq:Projections}), the crossing symmetry transformation in Eq.~(\ref{eq:MasterCrossing}), and remembering that the imaginary part of the physical 
scattering amplitude $\mathcal{A}_{\rm IJ}(s)$ 
in the s-channel is defined according to
\begin{equation}
\Im \mathcal{A}_{\rm IJ}(s) =\frac{1}{2i}\lim_{\epsilon\to 0^+}[
 \mathcal{A}_{\rm IJ}(s+i\epsilon) - \mathcal{A}_{\rm IJ}(s-i\epsilon) ]~,
\end{equation}
the contribution from the cuts in Eq.~(\ref{eq:CUTS}) can be rewritten as follows
\begin{eqnarray}
&&\int_{{\rm I-IV}}\frac{ds}{2\pi i}\frac{\mathcal{A}(s)}{s^2} =
\int_{0}^{\infty}\frac{ds}{4\pi s^2}\times
\nonumber\\ &&
\left\{(3\kappa_{00}+\kappa_{10}+\kappa_{01}-\kappa_{11})
[\Im\mathcal{A}_{00}(s)- 3\Im\mathcal{A}_{11}(s)]+\right.\nonumber \\
&&\hspace{2mm}(3\kappa_{00}+5\kappa_{10}-3\kappa_{01}-\kappa_{11})\Im\mathcal{A}_{10}(s)+\nonumber \\
&&\hspace{1.7mm}\left.(3\kappa_{00}-2\kappa_{10}+5\kappa_{01}-\kappa_{11})\Im\mathcal{A}_{01}(s)\right\}~.
\end{eqnarray}
Assuming left-right symmetry, i.e. taking $\Im\mathcal{A}_{10}(s)=\Im\mathcal{A}_{01}(s)\equiv 
\Im\mathcal{A}_{{\rm LR}}(s)$, we find
\begin{eqnarray}\label{eq:MASTER2}
&&
\int_{{\rm I-IV}}\frac{ds}{2\pi i}\frac{\mathcal{A}(s)}{s^2} =
(3\kappa_{00}+\kappa_{10}+\kappa_{01}-\kappa_{11})\times\nonumber \\ &&
\int_{0}^{\infty}\frac{ds}{4\pi s^2} [\Im\mathcal{A}_{00}(s) +2\Im\mathcal{A}_{{\rm LR}}(s)
-3\Im\mathcal{A}_{11}(s)]~.\nonumber \\
\end{eqnarray}
Notice that this symmetry is formally defined as the invariance under the exchange of the generators of $SU(2)_{\rm L}$
and $SU(2)_{\rm R}$, and it was introduced in Ref.~\cite{Agashe:2006at} to prevent the presence of  large corrections affecting the $Zb\bar{b}$ vertex.

Let us now consider the contribution from the big circle at infinity in Eq.~(\ref{eq:Separation}).
 The rule
of thumb in this kind of computation is to show that the scattering amplitude 
falls to zero, or at least remains constant, as $|s|\to \infty$; in this case, in fact, 
it is straightforward to see that
 the integral goes to zero as soon as the big circle is pushed to infinity.
 In
order to evaluate the integral following this criterium, we can retrace the argument already 
used in Ref.~\cite{Falkowski:2012vh}, and based on the application of the Regge theory.
Therefore, let us first try to recap in a nutshell the main prerogatives of this theory. 
Considering in full generality the process $ab\to cd$,
 the Regge theory reconstructs the behavior of the corresponding
  scattering amplitude $\mathcal{A}_{ab\to cd}(s,t)$ in the kinematical region $s\gg |t|$ according to
  the following expression \cite{Gribov:2009zz}
  \begin{equation}\label{eq:Reggeon}
  \mathcal{A}_{ab\to cd}(s,t) 
  \stackrel{s\gg |t|}{\sim} 
 Z~\gamma_{ac}(t)\gamma_{bd}(t)~
  s^{\alpha(t)}~,
  \end{equation}
  where $Z$ is a complex constant; Eq.~(\ref{eq:Reggeon}) can be interpreted considering the exchange
 of an object (the so-called Reggeon) with couplings $\gamma_{ac}(t)$, $\gamma_{bd}(t)$, and 
  angular momentum $\alpha(t)=\alpha(0)+\alpha^{\prime} t$; in the forward limit we have 
  \begin{equation}\label{eq:Reggeone}
    \mathcal{A}_{ab\to cd}(s,0) 
  \stackrel{s\to \infty}{\sim} 
 Z~\gamma_{ac}(0)\gamma_{bd}(0)~
  s^{\alpha(0)}~.
  \end{equation}
  On the other hand analyticity and unitarity, by virtue of the Froissart-Martin bound \cite{Froissart:1961ux,Martin:1962rt,Martin:1965jj}, impose the constraint\footnote{Notice that the Froissart-Martin bound controls the high-energy behavior of elastic scattering amplitude, i.e. 
  $\mathcal{A}_{ab \to ab}(s, t)$. However, using unitarity, it can be generalized also to inelastic scattering amplitude as in Eq.~(\ref{eq:FMGeneralizedBound}). We provide a proof of this generalization in Appendix~\ref{App:FMBound}.
  }
  \begin{equation}\label{eq:FMGeneralizedBound}
|\mathcal{A}_{ab \to cd}(s, 0)| \leqslant {\rm const}~s(\ln s)^2~.
  \end{equation}
  Regge amplitudes saturating the Froissart bound, therefore, grow like $s$, and give non-zero contribution 
  to the integral over $\mathcal{C}_{\infty}$. 
  According to Regge theory, this happens in correspondence of the exchange
  of a Reggeon with intercept $\alpha(t)=1$, and quantum number of the vacuum (i.e. without exchange 
  of isospin and charge, $\gamma_{ab}=C\delta_{ab}$). This trajectory is the
   Pomeron, and the corresponding amplitude reads
 \begin{equation}\label{eq:Pomerone}
    \mathcal{A}_{ab\to cd}(s,0) 
  \stackrel{s\to \infty}{\sim} 
 Z~C^2\delta_{ac}\delta_{bd}~
  s~.
  \end{equation}
 From Eqs.~(\ref{eq:A11}-\ref{eq:A01}) it follows that all the projections $\mathcal{A}_{\rm IJ}(s)$ can accommodate 
 the Pomeron exchange, and in particular we find
 \begin{equation}\label{eq:CInfinityContribution}
 \mathcal{A}(s)\sim Z~C^2(\kappa_{00}+
 \kappa_{10}+\kappa_{01}+\kappa_{11})s~.
 \end{equation}
 As in Ref.~\cite{Falkowski:2012vh}, we can exploit the freedom in the choice of the coefficients 
 $\kappa_{{\rm IJ}}$ in such a way that  $(\kappa_{00}+
 \kappa_{10}+\kappa_{01}+\kappa_{11})=0$, 
 $(3\kappa_{00}+ \kappa_{10}+\kappa_{01}-\kappa_{11})\neq 0$; in this case the contribution to the integral over the big circle at infinity
  originating from Eq.~(\ref{eq:CInfinityContribution}) vanishes.
  
 Combining Eq.~(\ref{eq:MASTER1}) and Eq.~(\ref{eq:MASTER2}) the final result is

 \begin{equation}\label{eq:SUMRULE}
 c_H=\frac{f^2}{4\pi}\int_{0}^{\infty}\frac{ds}{s}
 \left[
 \sigma^{\rm tot}_{00}(s) + 2\sigma_{\rm LR}^{\rm tot}(s)
 -3\sigma_{11}^{\rm tot}(s)
 \right]~,
 \end{equation}
where we made use of the optical theorem $s\sigma^{\rm tot}_{\rm IJ}(s) = \Im\mathcal{A}_{\rm IJ}(s)$,
being $\sigma^{\rm tot}_{\rm IJ}(s)$ the total cross section for the process ${\rm IJ}\to anything$.
The equality in Eq.~(\ref{eq:SUMRULE}) holds in the limit of vanishing gauge couplings $g_{\rm L},g_{\rm Y}\to 0$, and in the limit
of unbroken electroweak symmetry, where the relevant 
global symmetry governing the scattering amplitude is 
$SU(2)_{\rm L}\otimes SU(2)_{\rm R}$.
 Relaxing
the left-right symmetric condition, i.e. considering $\mathcal{A}_{10}\neq \mathcal{A}_{01}$, 
we can choose  the coefficient in such a way that $\kappa_{10}=\kappa_{01}=0$, 
$\kappa_{00}+\kappa_{11}=0$, $3\kappa_{00}-\kappa_{11}\neq0$. 
In this case we find the following generalization
  \begin{equation}\label{eq:SUMRULE2}
 c_H=\frac{f^2}{4\pi}\int_{0}^{\infty}\frac{ds}{s}
 \left[
 \sigma^{\rm tot}_{00}(s) + \sigma_{\rm 10}^{\rm tot}(s)+ \sigma_{\rm 01}^{\rm tot}(s)
 -3\sigma_{11}^{\rm tot}(s)
 \right]~.
 \end{equation}

\section{Discussion and Outlook}\label{sec:Discussion}

Let us now discuss some consequences of the sum rule derived in the last Section.
The sum rule connects the IR limit of the theory, represented by the coefficient $c_H$ of the 
effective 
SILH Lagrangian, with a combination of total cross sections that is valid, in principle, up to arbitrary high energy. 
This connection is completely general, because it does not rely on specific details -- apart from the postulates of  unitarity and 
analyticity -- of the underlying strong dynamics that, as a consequence, remains unknown.

One of the most intriguing consequences of this kind of sum rule,
as already emphasized in Refs.~\cite{Adams:2006sv,Low:2009di,Falkowski:2012vh}, 
is the possibility to investigate the positivity of $c_H$. The sign of this coefficient is particularly important both from a phenomenological and a theoretical viewpoint. On the one hand $c_H$ contributes to the Higgs propagator; as a consequence it modifies universally all the SM Higgs couplings, thus providing a 
direct connection with the corresponding 
measurements under investigation at the LHC \cite{Aad:2013wqa,CMS,LHCHiggsCrossSectionWorkingGroup:2012nn}. Considering the Higgs couplings with electroweak gauge bosons ($g_{hVV}$) and fermions
($g_{h\bar{f}f}$)  
 at the first order in $\xi \equiv v^2/f^2$ one finds \cite{Giudice:2007fh}
\begin{eqnarray}
k_{hVV} &=& 1-\frac{c_H}{2}\xi~,\label{eq:EWScaling}\\
k_{h\bar{f}f} &=& 1-\xi\left(\frac{c_H}{2}+c_f\right)~,\label{eq:EWferScaling}
\end{eqnarray}
in which we have defined the scaling factors $k_{i}\equiv g_i/g_i^{\rm SM}$, and where $c_y$ is the coefficient of the dim-6 operator
\begin{equation}
\mathcal{O}_f \equiv \frac{c_f y_f}{f^2}(H^{\dag}H)(\bar{f}_LH)f_R + {\rm h.c.}~,
\end{equation}
being $y_f$ the SM Yukawa coupling $y_f=\sqrt{2}m_f/v$. The sign of $c_H$, therefore, is crucial 
to understand if these deformations point towards a depletion or an enhancement 
of the Higgs couplings w.r.t. the SM predictions. 
\begin{figure}[!htb!]
  \includegraphics[width=0.85 \linewidth]{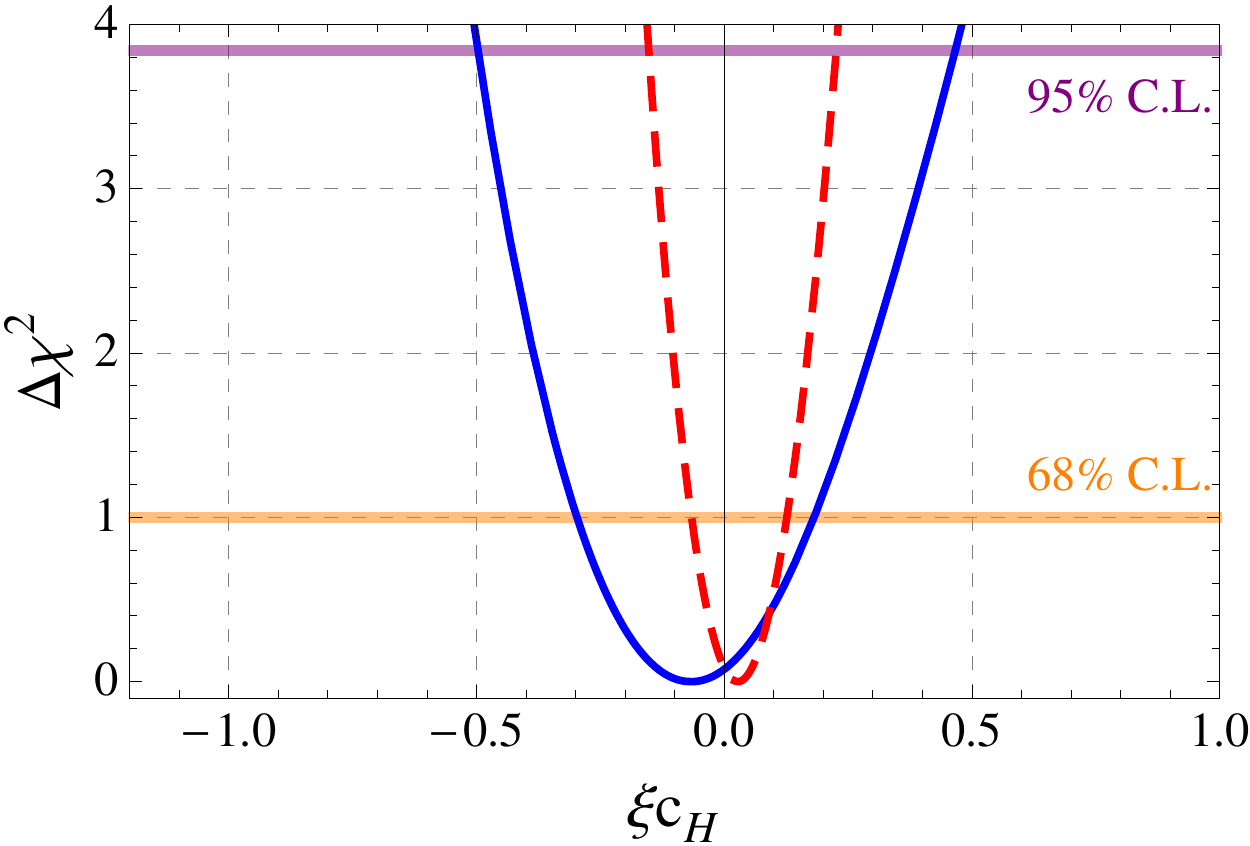}
 \caption{\textit{$\chi^2-\chi^2_{\rm min}$
 as a function of $\xi c_H$ obtained from  a fit to the Higgs data at the LHC. The blue solid line (red dashed line) is obtained marginalizing over $\xi c_{f=t,b,\tau}$ (setting $\xi c_{f=t,b,\tau}=0$).
  }}\label{fig:LHCFit}
\end{figure}
In Fig.~\ref{fig:LHCFit} we show the result of a chi-square fit of the Higgs data at the LHC (see Ref.~\cite{Falkowski:2013dza} for technical details), performed using as free parameters $\xi c_H$, $\xi c_{f=t,b,\tau}$ in Eqs.~(\ref{eq:EWScaling},\ref{eq:EWferScaling}). The blue solid line is obtained marginalizing over $\xi c_{f=t,b,\tau}$;
we find at 95\% C.L. $\xi c_H = -0.07^{+0.53}_{-0.43}$. For comparison, the one-dimensional fit obtained using $\xi c_H$ as free parameter and setting $\xi c_{f=t,b,\tau}=0$ gives at 
95\% C.L. $\xi c_H = 0.03^{+0.20}_{-0.18}$ (red dashed line).
The phenomenological relevance of $c_H$ immediately leads us to consider the theoretical implications of its sign.
To be more concrete, let us give some example. In Composite Higgs models based on a compact global symmetry $\mathcal{G}$, for instance, one always find a positive value
for $c_H$; in the Holographic Higgs model \cite{Agashe:2004rs}, e.g., we have $c_H=1$. Little Higgs models
also predicts a similar behavior; in the littlest Higgs model with 
custodial symmetry  \cite{Chang:2003zn}, e.g., one has $c_H = 1/2$. Composite Higgs models based on a non-compact 
global symmetry group $\mathcal{G}$, on the contrary, have negative $c_H$; in the minimal Composite Higgs model
based on $SO(4,1)/SO(4)$, e.g., one finds $c_H=-1$ (see Refs.~\cite{Falkowski:2012vh,RattazziTalk,LowTalk} and Appendix~\ref{app:Non-compact}), thus enhancing the Higgs coupling with the electroweak gauge bosons.

 The sum rule in Eq.~(\ref{eq:SUMRULE2}) can not fix the sign of $c_H$. On the right side, in fact, 
we have two combinations of total cross sections that
enter with opposite signs. Nevertheless, Eq.~(\ref{eq:SUMRULE2})
can isolate the source of negative contributions: they come from the total cross section 
in the channel with $(\textbf{3},\textbf{3})$ quantum numbers
under the global symmetry $SU(2)_{\rm L}\otimes SU(2)_{\rm R}$. 

This information contains  some interesting phenomenological
consequences. Following Ref.~\cite{Contino:2011np}, in fact,
one can assume that a resonance of the strong sector is accidentally 
lighter than the cut-off scale of the theory, $m_{\rho}\ll \Lambda\sim 4\pi f$. 
If so, this resonance may have sizable effects in the scattering processes involving the Higgs boson and/or the longitudinal gauge bosons $W^{\pm}_{\rm L}$, $Z_{\rm L}$. Using the Equivalence Theorem \cite{Chanowitz:1985hj} (for a recent discussion see Ref.~\cite{Wulzer:2013mza}) we can 
investigate these effects looking directly at the $\pi\otimes \pi^{\prime}$ scattering.
In this case the only resonances 
that can be exchanged are those that possess the correct quantum numbers according to the $SU(2)_{\rm L}\otimes SU(2)_{\rm R}$ decomposition in Eq.~(\ref{eq:SUDecomposition1}) 
\begin{eqnarray}
\eta&\sim&(\textbf{1},\textbf{1})~,~{\rm spin=0}~,~{\rm custodial=0}~,\\
\rho_{\rm L}&\sim&(\textbf{3},\textbf{1})~,~{\rm spin=1}~,~{\rm custodial=1}~,\\
\rho_{\rm R}&\sim&(\textbf{1},\textbf{3})~,~{\rm spin=1}~,~{\rm custodial=1}~,\\
\Delta &\sim&(\textbf{3},\textbf{3})~,~{\rm spin=0}~,~{\rm custodial=0+1+2}~,
\end{eqnarray}
where the spin assignment is dictated by Bose symmetry \cite{Contino:2011np}, and where we have indicated the decomposition under the custodial $SU(2)_{\rm C}$ group.
The role of these resonances in the $\pi\otimes \pi^{\prime}$ scattering can be immediately understood considering 
the expressions of the scattering amplitudes in terms of their $SU(2)_{\rm L}\otimes SU(2)_{\rm R}$
projections, collected in Eqs.~(\ref{eq:Amp1}-\ref{eq:Amp6}). More precisely,  we find the following classification (see Ref.~\cite{Contino:2011np} for the 
corresponding description based on the CCWZ effective Lagrangian \cite{Coleman:1969sm,Callan:1969sn}).
\begin{enumerate}[i)]

\item $\eta\sim(\textbf{1},\textbf{1})$. 

This resonance is left-right symmetric under  $SU(2)_{\rm L}\otimes SU(2)_{\rm R}$, and, therefore, it can not mediate 
left-right violating processes like $\pi^+\pi^-\to \pi^0 h$  in Eq.~(\ref{eq:Amp7}).
According to Eqs.~(\ref{eq:Amp4}-\ref{eq:Amp6}), moreover, $\eta$ is exchanged in the s-channel in the processes
$\pi^{\pm}\pi^{\mp}\to \pi^{\pm}\pi^{\mp}$, 
$\pi^{\pm}\pi^{\mp}\to \pi^0\pi^0$,
$\pi^{0}\pi^{0}\to \pi^0\pi^0$, as well as in the corresponding ones involving the Higgs boson
$\pi^{\pm}\pi^{\mp}\to hh$, $hh\to hh$, $\pi^0\pi^0\to hh$. Using the crossing symmetry transformation that relates the s- and the t-channel
(see Appendix~\ref{App:Crossing}), it follows that $\eta$ can be exchanged in the t-channel 
in the processes $\pi^{\pm}\pi^{\pm}\to \pi^{\pm}\pi^{\pm}$, $\pi^{\pm}\pi^{0}\to \pi^{\pm}\pi^{0}$, 
$\pi^{\pm}h\to \pi^{\pm}h$, $\pi^0h\to \pi^0h$. As pointed out in Ref.~\cite{Contino:2011np}, the t-channel exchange 
 results in a suppression of the cross section.

From the point of view of our sum rule, the existence of this resonance leads to an enhancement in
 the total cross section $\sigma_{00}^{\rm tot}(s)$; its presence, therefore, is favored in models featuring a positive value of $c_H$.

\item $\rho_{\rm L}\sim(\textbf{3},\textbf{1})$, $\rho_{\rm R}\sim(\textbf{1},\textbf{3})$. 

In order to preserve the left-right symmetry, both these resonances
 must be present with equal mass and couplings. In this case the amplitude describing left-right violating process
 like $\pi^{+}\pi^{-}\to \pi^0 h$ vanishes [see Eq.~(\ref{eq:Amp7})]. For definiteness, let us focus on the 
 case in which we have only $\rho_{\rm L}\sim(\textbf{3},\textbf{1})$. 
 According to Eqs.~(\ref{eq:Amp2},\ref{eq:Amp3},\ref{eq:Amp6}),  $\rho_{\rm L}$ 
 is exchanged in the s-channel in the processes $\pi^{\pm}\pi^{\mp}\to \pi^{\pm}\pi^{\mp}$, 
 $\pi^{\pm}\pi^0\to \pi^{\pm}\pi^0$, $\pi^{\pm}h\to \pi^{\pm}h$, thus enhancing the 
 corresponding cross sections. On the contrary, using again the crossing 
 symmetry, $\rho_{\rm L}$ can be exchanged in the t-channel in the processes 
 $\pi^{\pm}\pi^{\pm}\to \pi^{\pm}\pi^{\pm}$, $\pi^{\pm}\pi^{\mp}\to \pi^0\pi^0$,
 $\pi^{\pm}\pi^{\mp}\to hh$, suppressing the corresponding cross sections. Furthermore, as noticed in 
 \cite{Barbieri:2008cc,Contino:2011np} it turns out that the vector resonance is narrower w.r.t. the scalar one, thus implying a more promising scenario in Drell-Yan searches. 
 
 From the point of view of our sum rule, the existence of this resonance leads to an enhancement in
 the total cross section $\sigma_{10}^{\rm tot}(s)$; its presence, therefore, 
is favored in models featuring a positive value of $c_H$. Notice, moreover, that this kind of vector resonance 
is predicted in the minimal Composite Higgs model based on $SO(5)/SO(4)$ saturating the Weinberg
 sum rules \cite{Panico:2011pw,Matsedonskyi:2012ym,Marzocca:2012zn,Pomarol:2012qf,DeCurtis:2011yx,Redi:2012ha}.
 
\item $\Delta \sim(\textbf{3},\textbf{3})$.

This resonance is left-right symmetric under  $SU(2)_{\rm L}\otimes SU(2)_{\rm R}$, and, therefore, it can not mediate 
left-right violating processes like $\pi^+\pi^-\to \pi^0 h$   in Eq.~(\ref{eq:Amp7}). According to Eqs.~(\ref{eq:Amp1}-\ref{eq:Amp6}),  $\Delta$ 
 is exchanged in the s-channel in all the $\pi^a\pi^b\to \pi^c\pi^d$ processes, thus enhancing all the 
 corresponding cross sections.

From the point of view of our sum rule, the existence of this resonance leads to an enhancement in
 the total cross section $\sigma_{11}^{\rm tot}(s)$; its presence, therefore, is favored in models featuring a negative value of $c_H$.

\end{enumerate}

Finally, let us compare
 our sum rule with the existing literature. 
 Starting from Eq.~(\ref{eq:SUMRULE2}), using
 the optical theorem, and writing explicitly the scattering amplitudes 
 in terms of the charge eigenstates [see Appendix~\ref{eq:hhScattering}, 
 Eqs.~(\ref{eq:Amp1},\ref{eq:Amp6})] we find
 \begin{equation}
 c_{H} = \frac{f^2}{\pi}\int_{0}^{\infty}
 \frac{ds}{s}[\sigma_{+-}^{\rm tot}(s) - \sigma_{++}^{\rm tot}(s)]~;
 \end{equation}
 our sum rule, as a consequence, recovers the result obtained in Ref.~\cite{Low:2009di}. Similarly, 
 making use of the following
 $SU(2)_{\rm C}$ custodial decompositions under which the pions transform as a triplet \cite{Falkowski:2012vh}
 \begin{eqnarray}
 \mathcal{A}_{\pi^{\pm}\pi^{\pm}\to \pi^{\pm}\pi^{\pm}}(s) &=& \mathcal{T}_{2}(s)~,\\
 \mathcal{A}_{\pi^{\pm}\pi^{\mp}\to \pi^{\pm}\pi^{\mp}}(s)
  &=& \frac{\left[
  2\mathcal{T}_{0}(s) + 3\mathcal{T}_{1}(s) +\mathcal{T}_2(s)
  \right]}{6}~, 
 \end{eqnarray}
 where the eigenvalues $\mathcal{T}_{\rm I=0,1,2}$ are the analogous of the $\mathcal{A}_{\rm IJ}$ in Eq.~(\ref{eq:Projections}) but for the  $SU(2)_{\rm C}$
 combination $\textbf{3}\otimes \textbf{3}=\textbf{1}\oplus \textbf{3}\oplus \textbf{5}$, we find
 \begin{equation}
 c_H = \frac{f^2}{4}\int_{0}^{\infty}\frac{ds}{s^2}
 \left[
 \frac{1}{3}\Im\mathcal{T}_0(s)+\frac{1}{2}\Im\mathcal{T}_1(s) -\frac{5}{6}\Im\mathcal{T}_2(s)   
 \right]~,
 \end{equation}
 thus recovering the sum rule obtained in Ref.~\cite{Falkowski:2012vh}.
 
 Before concluding, a final caveat is mandatory.\footnote{We thank an anonymous referee for this comment.}
 The final result of this paper, Eq.~(\ref{eq:SUMRULE2}), surely provides useful indications 
 about the relation between possible deviations of the Higgs couplings 
 and the existence of  strongly coupled resonances. However, the statement 
 that a light resonance in a given channel ${\rm IJ}$ would enhance the corresponding contribution to $c_{H}$
 has to be taken with a grain of salt. What matters in the computation of the integral in Eq.~(\ref{eq:SUMRULE2}), in fact,
 is the ratio $\Gamma_{\rm IJ}^{\pi\pi}/M_{\rm IJ}^3\sim g_*^2/M_{\rm IJ}^2$, 
 where $\Gamma_{\rm IJ}^{\pi\pi}$ is the width of the IJ resonance into $\pi\pi$, $M_{\rm IJ}$ its mass, and 
 $g_*$ its coupling with the Nambu-Goldstone bosons. If a light resonance is more weakly coupled
 than a heavy one, then the contribution of the former does not dominate.

 \section{Conclusions}\label{sec:Conclusion}
 
 In this paper we have derived in the context of the SILH Lagrangian the following sum rule 
   \begin{equation}\label{eq:SUMRULE3}
 c_H=\frac{f^2}{4\pi}\int_{0}^{\infty}\frac{ds}{s}
 \left[
 \sigma^{\rm tot}_{00}(s) + \sigma_{\rm 10}^{\rm tot}(s)+ \sigma_{\rm 01}^{\rm tot}(s)
 -3\sigma_{11}^{\rm tot}(s)
 \right]~.
 \end{equation}
 The derivation of the sum rule is based on the axiomatic properties of Lorentz invariance, 
analyticity and unitarity, and it relies on the underlying 
 global symmetry $SU(2)_{\rm L}\otimes SU(2)_{\rm R}$.
The sum rule connects the low-energy coefficient $c_H$ to the UV properties of the theory,
 encoded into the
 combination of 
 total cross sections that appears on the right-hand side. The value of this coefficient is currently under experimental scrutiny at the LHC, and the possibility to extract from this measurement useful informations about the ultimate structure of the theory responsible for the electroweak symmetry breaking is of vital importance.
 
 For a given model featuring a SILH, the sum rule can give
 some useful insight about the corresponding UV-completion. In particular, the role of the resonances in the 
 scattering processes between longitudinal gauge bosons and/or the Higgs boson has been discussed.
 
 The sum rule favors the existence of a scalar resonance $\eta\sim(\textbf{1},\textbf{1})$ or a vector resonance
 $\rho_{\rm L}\sim (\textbf{3},\textbf{1})$ [or, equivalently, $\rho_{\rm R}\sim (\textbf{1},\textbf{3})$] in models with a positive value of $c_H$, like in Composite Higgs models based on a compact global symmetry group. 
 In presence of the scalar resonance $\eta\sim(\textbf{1},\textbf{1})$, in particular, the process $W_{\rm L}^{\pm}W_{\rm L}^{\mp}\to W_{\rm L}^{\pm}W_{\rm L}^{\mp}$ and the double Higgs production $W_{\rm L}^{\pm}W_{\rm L}^{\mp}\to hh$ and $Z_{\rm L}Z_{\rm L}\to hh$ are supposed to be enhanced.  
 In models featuring a negative value of $c_H$, on the contrary, the presence of a large contribution from a scalar 
 resonance $\Delta\sim (\textbf{3},\textbf{3})$ is mandatory.

\vspace{.5 cm}
{\bf Acknowledgments.} 
The author is beholden to Daniele Amati and Yuri Dokshitzer for enlightening discussions, and to Slava Rychkov for reading the manuscript and encouragments. The author is also grateful to Francesco Riva for important comments.
This work is supported by the ERC Advanced Grant n$^{\circ}$ $267985$, ``Electroweak Symmetry Breaking, Flavour and Dark Matter: One Solution for Three Mysteries" (DaMeSyFla).

\appendix

\section{Scattering amplitudes and the S-matrix}\label{app:Smatrix}
 
The remarkable goal of the \textit{S-matrix program}, developed during the sixties before the rise of QCD, was to 
construct and compute scattering amplitudes using only three postulates as guiding principles \cite{Gribov:2009zz}.
To be more concrete, 
let us consider in full generality the scattering $i\to f$ from an initial state $i$ to a final state $f$; the corresponding S-matrix element is
\begin{equation}\label{eq:SMatrix}
S_{fi}\equiv \langle f|S|i\rangle =
\delta_{fi}+i(2\pi)^4\delta^4(p_f - p_i)\mathcal{A}_{i\to f}~,
\end{equation}
where $\mathcal{A}_{i\to f}$ is the relativistic scattering amplitude. The aforementioned three principles are the following. 

\begin{enumerate}[i)]

\item The S-matrix is Lorentz invariant. This means that the scattering amplitude 
can be written as a function of the Lorentz invariants -- scalar products and rest masses -- involved in the process. Considering for definiteness the two-to-two scattering 
$ab \to cd$, these Lorentz invariants can be recast in terms of the usual Mandelstam variables
\begin{eqnarray}
s&\equiv& (p_a + p_b)^2=(p_a + p_c)^2~,\\
t&\equiv& (p_a - p_c)^2=(p_b - p_d)^2~,\\
u&\equiv& (p_a - p_d)^2=(p_b - p_c)^2~,
\end{eqnarray}
related by $s+t+u=\sum_{i=a,\dots d} m_i^2$. We denote the corresponding scattering amplitude as 
\begin{equation}\label{eq:MasterAmplitude}
\mathcal{A}_{ab\to cd}(s,t,u)~,
\end{equation}
bearing in mind, however, that $u$ is not an independent variable. In the following, whenever it is not necessary, we will omit the u-dependence.

\item The S-matrix is unitary, $S^{\dag}S=SS^{\dag}=\textbf{1}$. This property is a consequence of the conservation of  probability. In terms of the scattering amplitude in Eq.~(\ref{eq:SMatrix}) the unitarity condition reads
\begin{equation}\label{eq:MasterUnitary}
2 \Im \mathcal{A}_{i\to f} =  \sum_n \int d\Pi_n \mathcal{A}^*_{f\to n}\mathcal{A}_{i \to n}~,
\end{equation}
where $d\Pi_n$
is the $n$-particle phase-space measure.

\item The S-matrix is an analytical function of the Lorentz invariants regarded as variables in the complex plane. 
The singularities of the S-matrix are only those dictated by unitarity. It can be proved that this property is intimately connected with causality \cite{Gribov:2009zz}. Apart from the usual formalities,
the analyticity of a scattering amplitude finds an operative definition in the Cauchy integral
 formula
 \begin{equation}\label{eq:MasterAnalyticity}
 \mathcal{A}_{i\to f}(s,t)=\frac{1}{2\pi i}\int_{\mathcal{C}}
 ds^{\prime}\frac{ \mathcal{A}_{i\to f}(s^{\prime},t)}{s^{\prime} - s}~,
 \end{equation}
where $\mathcal{C}$ is a contour that does not enclose the singularities of $ \mathcal{A}_{i\to 
f}(s,t)$. Eqs.~(\ref{eq:MasterUnitary}, \ref{eq:MasterAnalyticity}) are the key equations of the S-matrix program:
 once the imaginary part of $\mathcal{A}_{i\to f}(s,t)$ is known, in fact, 
the Cauchy integral formula -- rewritten in terms of a dispersion relation \cite{Gribov:2009zz} -- allows to fully reconstruct the scattering amplitude.

\end{enumerate}


\section{The scattering amplitude $\mathcal{A}_{\pi^a\pi^b\to \pi^c\pi^d}$
}\label{eq:hhScattering}

In this Appendix we construct explicitly  the scattering amplitudes for the process
$\pi^{a}(p_1) + \pi^{b}(p_2) \to \pi^c(p_3) + \pi^d(p_4)$,
where $\pi^{a}=\pi^{\pm}, \pi^0, h$. In \ref{App:Symmetry}
we show how the $SU(2)_{\rm L}\otimes SU(2)_{\rm R}$
symmetry dictates the general structure of these amplitudes, 
while in \ref{App:Crossing} we discuss the corresponding transformations
of crossing symmetry. 

\subsection{The role of the $SU(2)_{\rm L}\otimes SU(2)_{\rm R}$
symmetry}\label{App:Symmetry}

In order to construct the scattering amplitude for the process
$\pi^{a}\pi^{b} \to \pi^c\pi^d$
 we make use of the symmetry $SU(2)_{\rm L}\otimes SU(2)_{\rm R}$ under which the Goldstone bosons and the Higgs boson transform according to the bi-doublet representation 
\begin{equation}\label{eq:DoubletAntidoublet}
H^{c}H=\left(
\begin{array}{cc}
  \frac{(h-i\pi^0) }{\sqrt{2}}&  \pi^+  \\
  -\pi^- & \frac{(h+i\pi^0) }{\sqrt{2}}
\end{array}
\right)\sim(\textbf{2},\textbf{2})_{SU(2)_{\rm L}\otimes SU(2)_{\rm R}}~,
\end{equation}
where $H^c\equiv i\sigma_2 H^*$ and
\begin{equation}
H^{c}H \stackrel{SU(2)_{\rm L}\otimes SU(2)_{\rm R}}{\longrightarrow} 
g_{\rm L}~H^{c}H~g_{\rm R}^{\dag}~.
\end{equation} 
From the composition of angular momenta it follows that 
the $\pi\otimes \pi^{\prime}$ combination admits the following deconstruction
\begin{equation}\label{eq:SUDecomposition}
(\textbf{2},\textbf{2})\otimes (\textbf{2},\textbf{2})
=(\textbf{1},\textbf{1})
\oplus (\textbf{3},\textbf{1})\oplus (\textbf{1},\textbf{3})\oplus (\textbf{3},\textbf{3})~;
\end{equation}
this means that we can organize the initial and the final state of the scattering process according to their 
$SU(2)_{\rm L}\otimes SU(2)_{\rm R}$ quantum numbers.
 To this purpose, we start 
from Eq.~(\ref{eq:DoubletAntidoublet}), labeling the states in the representation
$(\textbf{2},\textbf{2})$ using the notation $|t_{\rm L},t_{\rm L}^3;
t_{\rm R},t_{\rm R}^3\rangle$ 
\begin{eqnarray}
|\frac{1}{2}, \frac{1}{2}; \frac{1}{2}, \frac{1}{2}\rangle &=& |\pi^+\rangle\equiv |1\rangle~,\\
|\frac{1}{2}, -\frac{1}{2}; \frac{1}{2}, \frac{1}{2}\rangle &=&  \frac{1}{\sqrt{2}}(| h \rangle + i| \pi^0 \rangle )\equiv |2\rangle~,\\
|\frac{1}{2}, \frac{1}{2}; \frac{1}{2}, -\frac{1}{2}\rangle &=&  \frac{1}{\sqrt{2}}(| h \rangle - i| \pi^0 \rangle )\equiv |\bar{1}\rangle~,\\
|\frac{1}{2}, -\frac{1}{2}; \frac{1}{2}, -\frac{1}{2}\rangle &=& -|\pi^-\rangle\equiv |\bar{2}\rangle~,
\end{eqnarray}
 where the last definition is nothing but a shorthand notation. In the combination $\pi \otimes \pi^{\prime}$
 the states coming from the sum $\vec{T}_{\rm L} = \vec{t}_{\rm L} + \vec{t}_{\rm L}^{~\prime}$, 
 $\vec{T}_{\rm R} = \vec{t}_{\rm R} + \vec{t}_{\rm R}^{~\prime}$ generate the representation $(\textbf{2},\textbf{2})\otimes (\textbf{2},\textbf{2})$
 that we label as $|T_{\rm L}, T_{\rm L}^3; T_{\rm R}, T_{\rm R}^3 \rangle$. According to  Eq.~(\ref{eq:SUDecomposition}) we find \cite{Ciafaloni:2001vu,Ciafaloni:2009mm}
 \begin{itemize}

\item Singlet $(\textbf{1},\textbf{1})$, $(T_{\rm L}=0, T_{\rm R}=0)$
\begin{equation}\label{eq:Singlet}
|0,0;0,0\rangle = \frac{1}{2}(
|1\bar{2}\rangle - |\bar{1}2\rangle - |2\bar{1}\rangle + |\bar{2}1\rangle)
\end{equation}

\item Left Triplet $(\textbf{3},\textbf{1})$, $(T_{\rm L}=1, T_{\rm R}=0)$
\begin{eqnarray}
|1,1;0,0\rangle &=& \frac{1}{\sqrt{2}}(
|1\bar{1}\rangle - |\bar{1}1\rangle
)~,\\
|1,0;0,0\rangle &=& \frac{1}{2}(
|1\bar{2}\rangle - |\bar{1}2\rangle + |2\bar{1}\rangle  - |\bar{2}1\rangle
)~,\\
|1,-1;0,0\rangle &=& \frac{1}{2}(
|2\bar{2}\rangle - |\bar{2}2\rangle)~,
\end{eqnarray}

\item Right Triplet $(\textbf{1},\textbf{3})$, $(T_{\rm L}=0, T_{\rm R}=1)$
\begin{eqnarray}
|0,0;1,1\rangle &=& \frac{1}{\sqrt{2}}(
|12\rangle - |21\rangle
)~,\\
|0,0;1,0\rangle &=& \frac{1}{2}(
|1\bar{2}\rangle + |\bar{1}2\rangle - |2\bar{1}\rangle  - |\bar{2}1\rangle
)~,\\
|0,0;1,-1\rangle &=& \frac{1}{2}(
|\bar{1}\bar{2}\rangle - |\bar{2}\bar{1}\rangle)~,
\end{eqnarray}

\item Left-Right Triplet $(\textbf{3},\textbf{3})$, $(T_{\rm L}=1, T_{\rm R}=1)$
\begin{eqnarray}
|1,1;1,1\rangle &=&  |11\rangle~,\\
|1,1;1,0\rangle &=& \frac{1}{\sqrt{2}}(
|1\bar{1}\rangle - |\bar{1}1\rangle
)~,\\
|1,1;1,-1\rangle &=& 
|\bar{1}\bar{1}\rangle~,\\
|1,0;1,1\rangle &=&  \frac{1}{\sqrt{2}}(
|12\rangle + |21\rangle
)~,\\
|1,0;1,0\rangle &=& \frac{1}{2}(
|1\bar{2}\rangle + |\bar{1}2\rangle + |2\bar{1}\rangle + |\bar{2}1\rangle
)~,\\
|1,0;1,-1\rangle &=& 
\frac{1}{\sqrt{2}}(
|\bar{1}\bar{2}\rangle + |\bar{2}\bar{1}\rangle
)~,\\
|1,-1;1,1\rangle &=&  |22\rangle~,\\
|1,-1;1,0\rangle &=& \frac{1}{\sqrt{2}}(
|2\bar{2}\rangle + |\bar{2}2\rangle
)~,\\
|1,-1;1,-1\rangle &=& 
|\bar{2}\bar{2}\rangle~.\label{eq:Triplet}
\end{eqnarray}
 
 \end{itemize}
 After reversing the system formed by the eigenstates in Eqs.~(\ref{eq:Singlet}-\ref{eq:Triplet}), it is possible to use the
 Wigner-Eckart theorem to rewrite the scattering amplitude $\langle \pi^c\pi^d|
 \mathcal{A}|
 \pi^a\pi^b
 \rangle$
 in terms of the following eigenamplitudes
 \begin{equation}
 \langle T_{\rm L}, T_{\rm L}^{3};T_{\rm R}, T_{\rm R}^{3}|\mathcal{A}|
 T_{\rm L}^{\prime}, T_{\rm L}^{3\,\prime};T_{\rm R}^{\prime}, T_{\rm R}^{3\,\prime}
 \rangle = \mathcal{A}_{T_{\rm L} T_{\rm R}}
 \delta_{T_{\rm L}T_{\rm L}^{\prime}}
  \delta_{T_{\rm R}T_{\rm R}^{\prime}}~.
 \end{equation}
 As a consequence, we 
 can recast the amplitude
  $\langle \pi^c\pi^d|
 \mathcal{A}|
 \pi^a\pi^b
 \rangle$ 
as a function of the four scattering eigenvalues $\mathcal{A}_{00}$,
 $\mathcal{A}_{10}$, $\mathcal{A}_{01}$, $\mathcal{A}_{11}$. Reintroducing the notation
 $\langle \pi^c\pi^d|
 \mathcal{A}|
 \pi^a\pi^b
 \rangle = \mathcal{A}_{\pi^a\pi^b\to \pi^c\pi^d}(s,t,u)$ we  find
 \begin{eqnarray}
 \mathcal{A}_{\pi^{\pm}\pi^{\pm}\to \pi^{\pm}\pi^{\pm}}(s,t,u) &=& \mathcal{A}_{11}~,\label{eq:Amp1}\\
 \mathcal{A}_{\pi^{\pm}\pi^{0}\to \pi^{\pm}\pi^{0}}(s,t,u) &=& \frac{1}{4}
 (\mathcal{A}_{10}+2\mathcal{A}_{11}+\mathcal{A}_{01})~,\nonumber \label{eq:Amp2} \\
 \\
  \mathcal{A}_{\pi^{\pm}\pi^{0}\to \pi^0\pi^{\pm}}(s,t,u) &=&
 -\frac{1}{4}
  (\mathcal{A}_{10}-2\mathcal{A}_{11}+\mathcal{A}_{01})~,\nonumber \label{eq:Amp3}\\
  \\
   \mathcal{A}_{\pi^{\pm}\pi^{\mp}\to \pi^0\pi^{0}}(s,t,u) &=&
   \frac{1}{4}(\mathcal{A}_{00}-\mathcal{A}_{11})~,\label{eq:Amp4}\\
    \mathcal{A}_{\pi^{0}\pi^{0}\to \pi^0\pi^{0}}(s,t,u) &=&
   \frac{1}{4}(\mathcal{A}_{00}+3\mathcal{A}_{11})~,\label{eq:Amp5}\\
       \mathcal{A}_{\pi^{0}\pi^{0}\to hh}(s,t,u) &=&
   \frac{1}{4}(\mathcal{A}_{00}-\mathcal{A}_{11})~,\label{eq:Amp55}\\
     \mathcal{A}_{\pi^{\pm}\pi^{\mp}\to \pi^{\pm}\pi^{\mp}}(s,t,u) &=&
     \frac{1}{4}(\mathcal{A}_{00}+\mathcal{A}_{10}+
     \mathcal{A}_{01}+ \mathcal{A}_{11})~,\nonumber \label{eq:Amp6}\\ 
 \end{eqnarray}
 where on the right side the same kinematical dependence $\mathcal{A}_{\rm IJ}=\mathcal{A}_{\rm IJ}(s,t,u)$
 is understood. 
 Moreover, Eqs.~(\ref{eq:Amp2}-\ref{eq:Amp5}) hold true 
 replacing $\pi^0$ with the Higgs boson, i.e. for instance $\mathcal{A}_{\pi^{\pm}\pi^{0}\to \pi^{\pm}\pi^{0}}(s,t,u) 
 =\mathcal{A}_{\pi^{\pm}h\to \pi^{\pm}h}(s,t,u) $ and 
 $\mathcal{A}_{\pi^{0}\pi^{0}\to \pi^0\pi^{0}}(s,t,u)= \mathcal{A}_{hh\to hh}(s,t,u)$. On the contrary, we find
 $\mathcal{A}_{hh\to \pi^0 h}(s,t,u)=\mathcal{A}_{\pi^0\pi^0\to h\pi^0}(s,t,u)=0$. 
  Finally, notice that 
 we need one more amplitude
 in order to disentangle the combination $\mathcal{A}_{10}+\mathcal{A}_{01}$
 in Eqs.~(\ref{eq:Amp2},\ref{eq:Amp3},\ref{eq:Amp6}); in particular we find
 \begin{equation}\label{eq:Amp7}
   \mathcal{A}_{\pi^{+}\pi^{-}\to h\pi^0}(s,t,u) = \frac{i}{4}(\mathcal{A}_{10}-
   \mathcal{A}_{01}
  )~.
 \end{equation}
 This scattering amplitude is different from zero only breaking the left-right symmetry,  $\mathcal{A}_{10}\neq  \mathcal{A}_{01}$.
 Including Eq.~(\ref{eq:Amp7}) the system in Eqs.~(\ref{eq:Amp1}-\ref{eq:Amp6})
  can be immediately reversed, 
  and a trivial computation leads to the following expressions
  \begin{eqnarray}
  \mathcal{A}_{11}
  &=& \mathcal{A}_{\pi^{\pm}\pi^{\pm}\to \pi^{\pm}\pi^{\pm}}(s,t,u)~,\label{eq:A11}\\
  \mathcal{A}_{00}
  &=& 4\mathcal{A}_{\pi^{\pm}\pi^{\mp}\to \pi^{0}\pi^{0}}(s,t,u) +
  \mathcal{A}_{\pi^{\pm}\pi^{\pm}\to \pi^{\pm}\pi^{\pm}}(s,t,u)~,\nonumber\\ \label{eq:A00}\\
  \mathcal{A}_{10}
  &=& \mathcal{A}_{\pi^{\pm}\pi^{\pm}\to \pi^{\pm}\pi^{\pm}}(s,t,u)\nonumber \\
  &&-2\left[
  \mathcal{A}_{\pi^{\pm}\pi^{0}\to \pi^{0}\pi^{\pm}}(s,t,u)
  +i\mathcal{A}_{\pi^{+}\pi^{-}\to h\pi^{0}}(s,t,u)
  \right]~,\nonumber \\
  \\
   \mathcal{A}_{01}
  &=&\mathcal{A}_{\pi^{\pm}\pi^{\pm}\to \pi^{\pm}\pi^{\pm}}(s,t,u)\nonumber \\
  &&-2\left[
  \mathcal{A}_{\pi^{\pm}\pi^{0}\to \pi^{0}\pi^{\pm}}(s,t,u)
  -i\mathcal{A}_{\pi^{+}\pi^{-}\to h\pi^{0}}(s,t,u)
  \right]~.\nonumber  \label{eq:A01}\\
    \end{eqnarray}
    
\subsection{$SU(2)_{\rm L}\otimes SU(2)_{\rm R}$ crossing symmetry}\label{App:Crossing}    

One of the most powerful consequence of analyticity is crossing symmetry. 
Starting from Eq.~(\ref{eq:MasterAmplitude}), 
and defining the corresponding crossed scattering amplitudes
\begin{eqnarray}
{\rm s-channel:}&~& a + b \to c + d~,\label{eq:Schannel}\\
{\rm t-channel:}&~& a + \bar{c} \to \bar{b} + d~,\label{eq:Tchannel}\\
{\rm u-channel:}&~& a + \bar{d} \to c + \bar{b}~,\label{eq:Uchannel}
\end{eqnarray}
crossing symmetry is formally defined by the following relations
\begin{eqnarray}\label{eq:CrossingRelations}
\mathcal{A}_{ab\to cd}(s,t,u) &=& \mathcal{A}_{a\bar{c}\to \bar{b}d}(t,s,u)~,\label{eq:CrossingRelations1}\\ 
\mathcal{A}_{ab\to cd}(s,t,u) &=& \mathcal{A}_{a\bar{d} \to c\bar{b}}(u,t,s)~,\label{eq:CrossingRelations2}
\end{eqnarray}
and corresponds to the fact that, thanks to the analytical continuation, 
it is possible to describe all the processes in Eqs.~(\ref{eq:Schannel}-\ref{eq:Uchannel})
with the same analytical function but interchanging the role of the Mandelstam variables \cite{Gribov:2009zz}. In our case, because of the $SU(2)_{\rm L}\otimes SU(2)_{\rm R}$
structure, the crossing relations in 
Eqs.~(\ref{eq:CrossingRelations1},\ref{eq:CrossingRelations2})
have the following matrix form 
\begin{eqnarray}
\vec{\mathcal{A}}(s,t,u)&=& C_{st}~\vec{\mathcal{A}}(t,s,u)~,\\
\vec{\mathcal{A}}(s,t,u)&=& C_{su}~\vec{\mathcal{A}}(u,t,s)~,
\end{eqnarray}
where $\vec{\mathcal{A}} \equiv 
(\mathcal{A}_{00},\mathcal{A}_{10},
\mathcal{A}_{01},\mathcal{A}_{11})^T$. Using Eqs.~(\ref{eq:Amp1}-\ref{eq:Amp6})
and Eqs.~(\ref{eq:CrossingRelations1}-\ref{eq:CrossingRelations2}) we find
\begin{equation}
C_{st} = \frac{1}{4}
\left(
\begin{array}{cccc}
 1 & 3  & 3 & 9 \\
 1 & -1  & 3 & -3 \\
 1 & 3  & -1  & -3 \\
 1 & -1 & -1 & 1
\end{array}
\right)~,
\end{equation}
\begin{equation}\label{eq:CrossingMatrixU}
C_{su} = \frac{1}{4}
\left(
\begin{array}{cccc}
 1 & -3  & -3 & 9 \\
 -1 & -1  & 3 & 3 \\
 -1 & 3  & -1  & 3 \\
 1 & 1 & 1 & 1
\end{array}
\right)~.
\end{equation}
As a simple cross-check, these matrices satisfy the relations $C_{st}=C_{st}^{-1}$, $C_{su}=C_{su}^{-1}$.

\section{On the generalization of the Froissart-Martin bound for inelastic  amplitudes}\label{App:FMBound}

The Froissart-Martin bound \cite{Froissart:1961ux,Martin:1962rt,Martin:1965jj} controls the behavior of elastic scattering amplitudes at high energies.\footnote{This bound has been obtained in Ref.~\cite{Froissart:1961ux,Martin:1962rt} assuming analyticity and unitarity, and further 
re-examined in Ref.~\cite{Martin:1965jj} using only analytic properties from axiomatic
 quantum field theory.} In particular, considering the scattering amplitude $\mathcal{A}_{ab\to ab}(s, \cos\theta)$ --  being $\theta$ the scattering angle in the c.o.m. frame, with $t=-2(s/4-m^2)(1-\cos\theta)$ -- we have for real $s\to \infty$
\begin{eqnarray}
|\mathcal{A}_{ab\to ab}(s, \cos\theta = 1)| &\leqslant& {\rm const}~s  (\ln s)^2~,\label{eq:FMbound}\\
|\mathcal{A}_{ab\to ab}(s, |\cos\theta| < 1)| &\leqslant&
 {\rm const}~\frac{s^{\frac{3}{4}}(\ln s)^{\frac{3}{2}}}{(\sin\theta)^{\frac{1}{2}}}~.
\end{eqnarray}
Using the optical theorem the first inequality can be immediately translated into a bound on the high-energy behavior 
of the total cross section $\sigma^{\rm tot}_{ab}(s)$ describing the process $ab \to anything$
\begin{equation}
\sigma^{\rm tot}_{ab}(s) \leqslant  {\rm const}~(\ln s)^2~.
\end{equation}
From a more general point of view, one can be interested in inelastic processes where initial 
and final state are different \cite{Logunov:1971ni,Logunov:1971nq}. 
In the following we shall derive a generalization of the Froissart-Martin bound in 
Eq.~(\ref{eq:FMbound}), and our proof goes as follows.

Let us start  considering the inelastic scattering process $ab \to cd$  among scalar particles with, respectively, four-momenta $p_1$, $p_2$, $p_3$, $p_4$. 
For simplicity we assume that all the masses are equal, $s+t+u=4m^2$, with $\sqrt{s}\gg m$. In this limit the differential cross section is
\begin{equation}
\frac{d\sigma_{ab\to cd}}{d\cos\theta} = \frac{1}{32\pi s}|\mathcal{A}_{
ab\to cd}(s,\cos\theta)|^2~,
\end{equation}
where $\mathcal{A}_{
ab\to cd}(s,\cos\theta)$ is the scattering amplitude. The total cross section for the process $ab \to cd$ is therefore
\begin{equation}\label{eq:TotalXSec}
\sigma_{ab\to cd}(s) = \frac{1}{32\pi s}\int_{-1}^{+1}d\cos\theta
|\mathcal{A}_{
ab\to cd}(s,\cos\theta)|^2~.
\end{equation}

On a general ground we can set the following chain of inequalities 
\begin{equation}
\sigma_{ab\to cd}(s) < \sigma^{\rm inelastic}_{ab \to cd, \dots} < 
\sigma^{\rm tot}_{ab} \leqslant  {\rm const}~(\ln s)^2~,
\end{equation}
and, as a consequence, we obtain 
\begin{equation}
\int_{-1}^{+1}d\cos\theta
|\mathcal{A}_{
ab\to cd}(s,\cos\theta)|^2 \leqslant {\rm const}~s (\ln s)^2~.
\end{equation}
However, given that we are interested in the forward limit of the scattering amplitude, $|\mathcal{A}_{
ab\to cd}(s,\cos\theta = 1)|$, this result is not enough for our purposes. In order to put a bound on $|\mathcal{A}_{
ab\to cd}(s,\cos\theta = 1)|$, we proceed following three steps.
 
$1.$~We introduce the partial wave expansion 
\begin{equation}
\mathcal{A}_{
ab\to cd}(s,\cos\theta) = \sum_{l=0}^{\infty}
(2l + 1)a_{l}(s)P_{l}(\cos\theta)~,
\end{equation}
where $a_l(s)$ are the partial wave amplitudes and $P_{l}(\cos\theta)$ the Legendre polynomials. 
Bearing in mind that $P_{l}(1) = 1$ we have for the forward scattering amplitude
\begin{equation}
\left.
\frac{d\sigma_{ab\to cd}}{d\cos\theta}
\right|_{\theta = 0} = \frac{1}{32\pi s}\sum_{l,m=0}^{\infty}
(2l+1)(2m+1)a_{l}(s)a_m^*(s)~,
\end{equation}
and, using the Cauchy-Bunyakovskii inequality,
\begin{equation}\label{eq:MasterAn}
\left.\frac{d\sigma_{ab\to cd}}{d\cos\theta}
\right|_{\theta = 0} \leqslant 
 \frac{1}{32\pi s}\sum_{l,m=0}^{\infty}
(2l+1)(2m+1)|a_{l}(s)||a_m(s)|~.
\end{equation}

On the other hand, considering the total cross section in Eq.~(\ref{eq:TotalXSec}), we have
\begin{equation}
\sigma_{ab\to cd}(s) = \frac{1}{32\pi s}\sum_{l=0}^{\infty}2(2l + 1)|a_l(s)|^2~,
\end{equation}
where we made use of the orthogonality relation 
\begin{equation}
\int_{-1}^{+1}dx P_{l}(x)P_{m}(x) = \frac{2\delta_{lm}}{(2l + 1)}~.
\end{equation}

$2.$~The next step is to relate the partial wave amplitudes $a_l(s)$ in Eq.~(\ref{eq:MasterAn})
to the amplitude describing the corresponding elastic process $ab \to ab$. 
To this purpose, we use the unitarity condition in Eq.~(\ref{eq:MasterUnitary});
 for a scattering process $1+2\to n$, with initial state 
$|p_1, p_2\rangle$ and final state $|p_{1}^{\prime},\dots,p_{n}^{\prime}\rangle$ we have
\begin{equation}
d\Pi_n = (2\pi)^4\delta^{(4)}(p_1 + p_2 - \sum_{j=1}^n p_j^{\prime})
\prod_{j=1}^n \frac{d^3\textbf{p}^{\prime}_j}{(2\pi)^3 2E_{j}^{\prime}}~.
\end{equation}
Considering the two-body elastic scattering process $ab\to ab$,
and writing explicitly the independent kinematical variables, Eq.~(\ref{eq:MasterUnitary}) becomes\footnote{Notice that 
$\mathcal{A}_{ab\to ab}$ is the transition amplitude for the scattering process $ab\to ab$ in which the direction of motion is unchanged (initial and final state are equal). In other words, we are dealing with the elastic amplitude describing the forward scattering.}
\begin{equation}
2 \Im \mathcal{A}_{ab\to ab}(s,\cos\theta =1) = 
\sum_n \int d\Pi_n |\mathcal{A}_{ab \to n}|^2~.
\end{equation}
The right-hand side is a sum of positive numbers. Extracting only the process $ab\to cd$, we have the following inequality
\begin{eqnarray}
2 \Im \mathcal{A}_{ab\to ab}(s, 1) &\geqslant& \int_{-1}^{+1}
\frac{d\cos\theta}{16\pi} |\mathcal{A}_{ab \to cd}(s,\cos\theta)|^2\nonumber \\
&=& \frac{1}{16\pi}
\sum_{l=0}^{\infty}2(2l + 1)|a_l(s)|^2~.
\end{eqnarray}
The elastic amplitude $ \mathcal{A}_{ab\to ab}(s, 1)$, in turn, can be expanded in partial waves 
\begin{equation}
\mathcal{A}_{ab\to ab}(s, 1) = \sum_{l=0}^{\infty}(2l + 1)f_l(s)~,
\end{equation}
leading to 
\begin{equation}\label{eq:OrFroi}
|f_l(s)| \geqslant \frac{1}{16\pi}|a_l(s)|^2~,
\end{equation}
being $|f_l(s)| \geqslant \Im f_l(s)$. This inequality connects the partial wave amplitudes describing the elastic process 
$ab \to ab$, $f_l(s)$, to those describing the forward inelastic process $ab \to cd$, $a_l(s)$, 
according to Eq.~(\ref{eq:MasterAn}). 

$3.$~Finally, combining Eq.~(\ref{eq:MasterAn}) and Eq.~(\ref{eq:OrFroi}), we are now in the position to use the original argument of the Froissart theorem. This argument
relies on the fact that, in the large $s,l$ limit, we have the asymptotic behavior 
\begin{equation}
f_l(s)_{l,s\to \infty} \sim  \exp\left[
-\left(\frac{2m}{\sqrt{s}}\right)l +\delta \ln s
\right]~,
\end{equation}
where $\delta$ is an integer. This simply means that the partial waves with $l \gtrsim c\sqrt{s}\ln s$, where $c$ is some constant,
can be neglected. Because of unitarity all the remaining ones, moreover, are bounded according to 
$|f_l(s)| \leqslant 16\pi$. All in all we find
\begin{eqnarray}
\left. \frac{d\sigma_{ab\to cd}}{d\cos\theta}
\right|_{\theta = 0}
&\leqslant& \frac{8\pi}{s}
\sum_{l=0}^{c\sqrt{s}\ln s}(2l + 1)\sum_{m=0}^{c\sqrt{s}\ln s}(2m + 1)\nonumber \\
&\simeq& {\rm const}~s(\ln s)^4~,
\end{eqnarray}
and the final result is
\begin{equation}
|\mathcal{A}_{ab \to cd}(s, \cos\theta = 1)| \leqslant {\rm const}~s(\ln s)^2~.
\end{equation}

\section{The non-linear $\sigma$-model based on $SO(4,1)/SO(4)$}\label{app:Non-compact}

In this Appendix, we construct the non-linear  $\sigma$-model based on the coset $SO(4,1)/SO(4)$ following the CCWZ prescription, originally proposed in Refs.~\cite{Coleman:1969sm,Callan:1969sn} 
considering compact, connected, semisimple Lie group $\mathcal{G}$. The correspondent generalization to the case 
in which $\mathcal{G}$ is a non-compact group, and $\mathcal{H}$ is its maximal compact subgroup [as in $SO(4,1)/SO(4)$] is known in the context of supergravity theories (see, e.g., Ref.~\cite{Gursey:1979tu,Gates:1984kr,Yilmaz:2003fp}).

The de Sitter group $SO(4,1)$ \cite{SO41} finds an intuitive realization as the ten-parameter group of transformation matrices that acting on the five variables $w$, $x$, $y$, $z$, $t$ holds invariant the indefinite quadratic form
\begin{equation}
-x^2-y^2-z^2-w^2+t^2~.
\end{equation}
The generators of the corresponding algebra $L_{ab}=-L_{ba}$ satisfy the commutation relations
\begin{equation}
\left[L_{ab},L_{cd}\right]= i\left(-g_{ac}L_{bd}+g_{ad}L_{bc}+g_{bc}L_{ad}-g_{bd}L_{ac}\right)~,
\end{equation}
where $g={\rm diag}(-1,-1,-1,-1,+1)$ is the internal metric of the algebra. The explicit solution used throughout this paper is 
\begin{equation}\label{eq:ExplGen}
(L_{ab})_{ij} = i\left(\delta_{ia}~g_{bj} - \delta_{ib}~g_{aj}\right)~.
\end{equation}
The maximal compact subgroup of $SO(4,1)$ is the special orthogonal group $SO(4)$. In more detail, recasting the generators as follows
\begin{equation}
L_{ab} = \epsilon_{abc} J_c~,~~~~L_{a4} = K_a~,~~~~a,b,c=1,2,3~,
\end{equation}
the algebra of the isomorphism $SO(4)\approx SU(2)_{\rm L}\otimes SU(2)_{\rm R}$ can be recovered defining
\begin{equation}\label{eq:TLTR}
\vec{T}_{\rm L}\equiv \frac{1}{2}\left(\vec{J} + \vec{K}\right)~,~~~~
\vec{T}_{\rm R}\equiv \frac{1}{2}\left(\vec{J} - \vec{K}\right)~,
\end{equation}
while the remaining generators $T_{\rm c}^{a=1,_{\cdots},4}\equiv L_{a5}/\sqrt{2}$ define the coset $SO(4,1)/SO(4)$. Notice that, using the explicit realization in Eq. (\ref{eq:ExplGen}), the generators of the de Sitter group are normalized as follows 
\begin{eqnarray}
{\rm Tr}\left(T_{\rm L}^{a}~T_{\rm L}^{b}\right) &=&\delta_{ab}~,\\
{\rm Tr}\left(T_{\rm R}^{a}~T_{\rm R}^{b}\right) &=&\delta_{ab}~,\\
-{\rm Tr}\left(T_{\rm c}^{a}~T_{\rm c}^{b}\right) &=&\delta_{ab}~.
\end{eqnarray}

The coset $SO(4,1)/SO(4)$ is the four-dimensional hyperbolic space $\mathcal{H}^4$. To describe this space, first we introduce the coordinates $\varphi_{1,_{\cdots},4}$ to parametrize the left cosets, then we define the coset representative field 
\begin{equation}
U(\vec{\varphi})=\exp\left[\frac{i\sqrt{2}}{f}\,\varphi_{a}\,T_{\rm c}^{a} \right]~.
\end{equation}
The coset representative field is an element of the group $SO(4,1)$ which transform under global $SO(4,1)$ transformation from the left and local $SO(4)$ transformation from the right.
It satisfies the defining relation of $SO(4,1)$, namely $U^T\,g\,U = g$, and its inverse can be build using the metric $g$ as $U^{-1}=g\,U\,g$. In the unitary gauge $\vec{\varphi}=(0,0,0,h)$ the coset representative field takes the explicit matrix form
\begin{equation}
U(h)=\left(
\begin{array}{ccccc}
1  & 0  & 0 & 0 & 0 \\
0  & 1  & 0 & 0 & 0\\
 0 & 0  & 1 & 0 & 0 \\
0  & 0  & 0 & \cosh\frac{h}{f} & -\sinh\frac{h}{f} \\
0  & 0  & 0 & -\sinh\frac{h}{f} & \cosh\frac{h}{f}
\end{array}
\right)~.
\end{equation}
Following Ref.~\cite{Gursey:1979tu,Gates:1984kr,Yilmaz:2003fp}, we  introduce the  decomposition
\begin{equation}\label{eq:CCWZ}
U^{-1}\left(\partial_{\mu}U\right)= d_{\mu}+E_{\mu}~,
\end{equation}
where $d_{\mu}\equiv d_{\mu}^{a}T_{\rm C}^{a}$, $E_{\mu}\equiv E_{{\rm L},\mu}^{a}T_{\rm L}^{a}+E_{{\rm R},\mu}^{a}T_{\rm R}^{a}$. Gauging the $SU(2)_{\rm L}\otimes U(1)_{\rm Y}$ SM subgroup amounts to promoting the ordinary derivatives to covariant ones, $\partial_{\mu}\to D_{\mu}=\partial_{\mu}-i(g_{\rm L}W_{\mu}^{a}T_{\rm L}^{a}+g_{\rm Y}B_{\mu}T_{\rm R}^{3})$. The gauged version of Eq. (\ref{eq:CCWZ})  becomes $U^{-1}\left(D_{\mu}U\right)= \hat{d}_{\mu}+\hat{E}_{\mu}$, with  $\hat{d}_{\mu}\equiv \hat{d}_{\mu}^{a}T_{\rm C}^{a}$, $\hat{E}_{\mu}\equiv \hat{E}_{{\rm L},\mu}^{a}T_{\rm L}^{a}+\hat{E}_{{\rm R},\mu}^{a}T_{\rm R}^{a}$. The leading order non linear $\sigma$-model Lagrangian for the SM gauge and Goldstone bosons is
\begin{equation}
\mathcal{L}_{\sigma}=-\frac{1}{4}W_{\mu\nu}^{a}W^{a,\mu\nu}-\frac{1}{4}B_{\mu\nu}B^{\mu\nu}
+\frac{f^2}{4}{\rm Tr}\left(\hat{d}_{\mu}\hat{d}^{\mu}\right)~.
\end{equation} 
We find
\begin{eqnarray}
\mathcal{L}_{\sigma}&=&-\frac{1}{4}W_{\mu\nu}^{a}W^{a,\mu\nu}-\frac{1}{4}B_{\mu\nu}B^{\mu\nu}
+\frac{1}{2}(\partial_{\mu}h)(\partial^{\mu}h)\nonumber\\
&&+\frac{1}{8}g_{\rm L}^2f^2(W_{\mu}^{1}W^{1,\mu}+W_{\mu}^{2}W^{2,\mu})\sinh^2\frac{h}{f}\nonumber \\ 
&&+
\frac{1}{8}f^2(g_{\rm L}W_{\mu}^3-g_{\rm Y}B_{\mu})(g_{\rm L}W^{3,\mu}-g_{\rm Y}
B^{\mu})\sinh^2\frac{h}{f}~.\nonumber\\
\end{eqnarray}
From the above Lagrangian we read the value of the electroweak gauge boson masses 
\begin{eqnarray}
m_{W}^2(\langle h\rangle)&=& \frac{1}{4}f^2g_{\rm L}^2\sinh^2\frac{\langle h\rangle}{f}=\frac{1}{4}g_{\rm L}^2v^2~,\\
m_{Z}^2(\langle h\rangle)&=& \frac{1}{4}f^2(g_{\rm L}^2+g_{\rm Y}^2)\sinh^2\frac{\langle h\rangle}{f}=\frac{1}{4}(g_{\rm L}^2+g_{\rm Y}^2)v^2~,\nonumber \\
\end{eqnarray}
from which we obtain
\begin{equation}
\sinh^2\frac{\langle h\rangle}{f} = \frac{v^2}{f^2}\equiv \xi~.
\end{equation}
Following Ref.~\cite{Giudice:2007fh}, the scaling factor describing the coupling of the Higgs with the 
electroweak gauge boson $V$ is
\begin{equation}\label{eq:NonCompact1}
k_{hVV}=\left.
\frac{1}{g_{\rm L}m_{V}(h)}
\frac{\partial m_V^2(h)}{\partial h}
\right|_{h=\langle h\rangle}=\sqrt{1+\xi}~.
\end{equation}
Notice that in Composite Higgs models based on a compact global symmetry group $\mathcal{G}$ one finds $k_{hVV}=\sqrt{1-\xi}$.
Similarly, the scaling factor describing the Higgs quadratic coupling to the electroweak gauge boson $V$ follows from
\begin{equation}\label{eq:NonCompact2}
k_{hhVV}=\left.
\frac{4}{g_{\rm L}^2}\frac{\partial^2 \mathcal{L}_{hhVV}}{\partial h^2}
\right|_{h=\langle h\rangle}=1+2\xi~,
\end{equation}
where $\mathcal{L}_{hhWW}= (f^2/8)g_{\rm L}^2\sinh^2(h/f)$, $\mathcal{L}_{hhBB}= (f^2/8)g_{\rm Y}^2\sinh^2(h/f)$.
In Composite Higgs models based on a compact  global symmetry group $\mathcal{G}$ 
one finds $k_{hhVV}=1-2\xi$.

\begin{figure}[!htb!]
  \includegraphics[width=0.85 \linewidth]{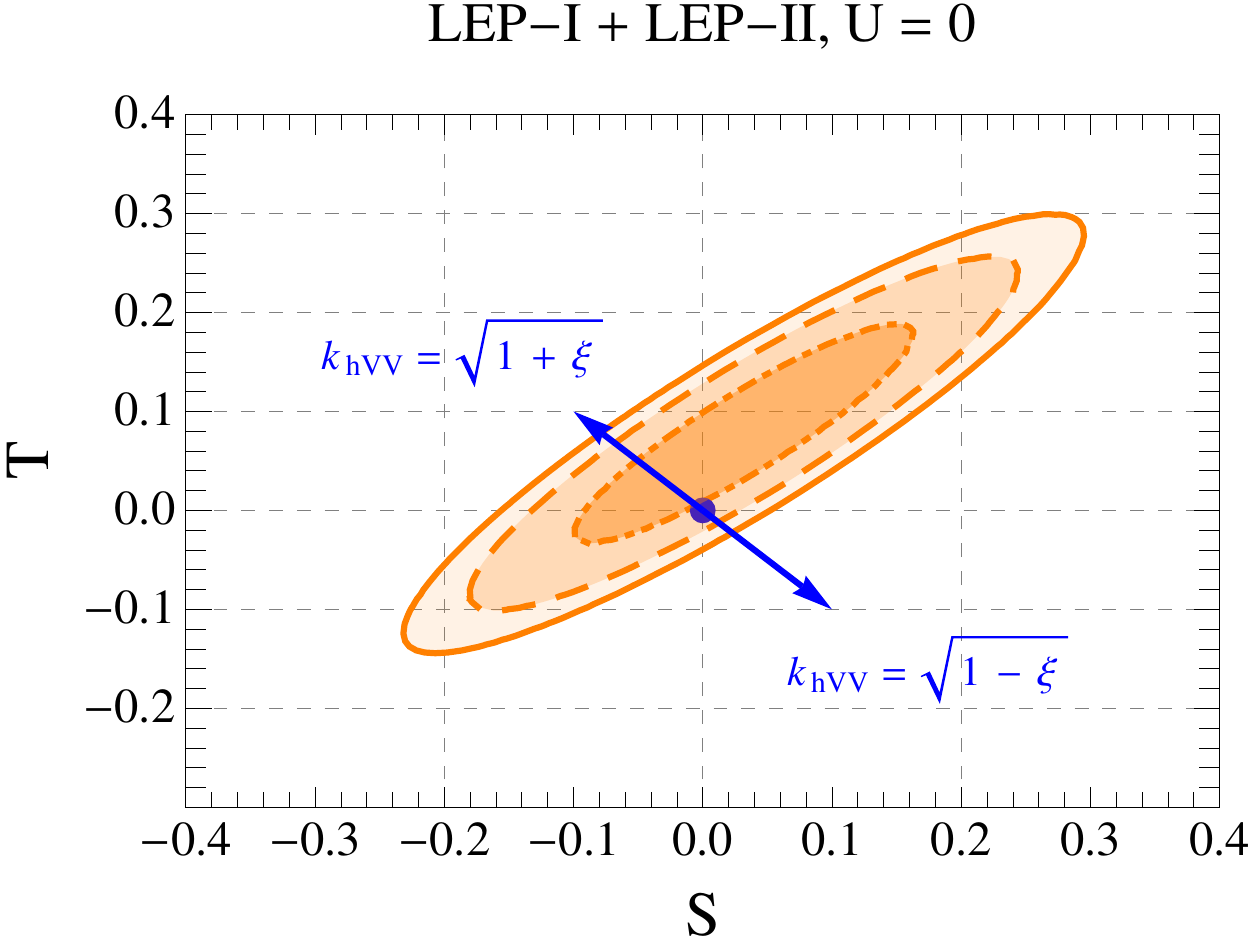}
 \caption{\textit{Confidence regions (68\%, 95\%, 99\% C.L.) for the $S$ and $T$ oblique parameters ($U=0$) obtained from the fit of  LEP-I and LEP-II data. 
  }}\label{fig:NonCompactFit}
\end{figure}

The flipped sign in Eq.~(\ref{eq:NonCompact1}) has important phenomenological implications \cite{Contino:2013gna}, summarized  for the sake of clarity in Fig.~\ref{fig:NonCompactFit}, where we fit the LEP data in the plane defined by the oblique parameters $S$ and $T$ \cite{Peskin:1991sw,Barbieri:2004qk} (see Ref.~\cite{Falkowski:2013dza,Ciuchini:2013pca} for a detailed discussion about the fit). The reference point at which $S$ and $T$ vanish is defined by the SM with $m_h = 126$ GeV and $m_t = 173.5$ GeV, and it lies on the boundary of the 68\% confidence contour, in agreement with the experimental data.
Deviations of the Higgs couplings with the electroweak gauge bosons $W^{\pm},Z$ w.r.t. their SM values generate logarithmic correction \cite{Barbieri:2007bh} to the $S$ and $T$ parameters in the directions shown by the representative arrows in the plot. The correction in Eq.~(\ref{eq:NonCompact1}) points towards the 
favored region, thus alleviating the tension with the electroweak precision measurements  that affects the Composite Higgs models based on a compact global symmetry \cite{Contino:2010rs}. It is important to keep in mind, however, 
that extra contributions coming from the strong sector can drastically modify this picture.

\end{document}